\documentclass[aps, pra, reprint, longbibliography,
superscriptaddress]{revtex4-1}

\usepackage{hyperref}
\usepackage{graphicx}
\usepackage[utf8]{inputenc}
\usepackage{xcolor}
\usepackage{amsmath, amssymb}
\usepackage{physics}
\usepackage{ulem}

\begin{document}

\preprint{APS/123-QED}

\title{Thermal behavior of Bose-Einstein condensates of polar molecules}

\author{J. S\'anchez-Baena}
\email{juan.sanchez.baena@upc.edu}
\affiliation{Departament de F\'isica, Universitat Polit\`ecnica de Catalunya, Campus Nord B4-B5, 08034 Barcelona, Spain}
\author{G. Pascual}
\affiliation{Departament de F\'isica, Universitat Polit\`ecnica de Catalunya, Campus Nord B4-B5, 08034 Barcelona, Spain}
\author{R. Bombín}
\affiliation{Université de Bordeaux, 351 Cours de la Libération, 33405 Talence, France}
\author{F. Mazzanti}
\affiliation{Departament de F\'isica, Universitat Polit\`ecnica de Catalunya, Campus Nord B4-B5, 08034 Barcelona, Spain}
\author{J. Boronat}
\affiliation{Departament de F\'isica, Universitat Polit\`ecnica de Catalunya, Campus Nord B4-B5, 08034 Barcelona, Spain}

\date{\today}

\begin{abstract}

We use the finite-temperature extended Gross-Pitaevskii equation (TeGPE) to
study a condensate of dipolar NaCs molecules under the conditions of the very 
recent, breakthrough experiment [Bigagli \textit{et al.}, Nature \textbf{631}, 
289 (2024)]. We report the condensate fraction of the system, and its density 
profile after a time-of flight expansion for the coldest experimental case, 
finding excellent agreement with the experimental measurements.  We also report 
the peak density of the ground state and establish a comparison with
the experimental estimates.
Our results, derived from the  TeGPE formalism, successfully describe the Bose-Einstein condensation of polar molecules at finite temperature.

\end{abstract}

\maketitle

\section{\label{sec:introduction}Introduction}

The recent progress in the control and reduction of reactive losses for 
ultracold dipolar 
molecules~\cite{Gorshkov08,Junyu2023,Karam2023,Bigagli2023,Mukherjee2024} has 
culminated in the first realization of a Bose-Einstein condensate (BEC) of 
dipolar 
molecules~\cite{Bigagli2024}, which represents a major breakthrough in the 
field. While the condensate still lies in the dilute regime, it represents 
the first step towards the realization of highly correlated, dipole-dominated
molecular systems~\cite{langen_prl}. In fact, 
the physics of ultracold quantum dipolar atomic gases has been the subject of 
intense experimental and
theoretical activity in the last decade. This is due to the unique combination 
of traits of the dipole-dipole interaction (DDI), mainly its anisotropy and  
long-range behavior. Due to the high controllability achievable with ultracold
quantum gases, dipolar systems represent the perfect testbed to study the wide 
variety of physical phenomena that arises thanks to their unique properties. 
This includes, most notably, the emergence of ultradilute liquid dipolar 
droplets with magnetic atoms~\cite{Pfau:nature:2016,Pfau:nature2:2016,Pfau:PRL:2016,ferlaino16,
bottcher19}, 
supersolids~\cite{Modugno:PRL:2019,Pfau:PRX:2019,Ferlaino:PRX:2019,
Tanzi:Nature:2019,Guo:Nature:2019,Tanzi:Science:2021,norcia21:nature,
BiagioniPRX2022,Ferlaino:PRL:2021} or anomalous thermal 
behavior~\cite{Ferlaino:PRL:2021,baena22,baena24,He2024}, to name a few. At present, the vast majority of experiments on dipolar gases have been realized
using highly magnetic atoms, like Dysprosium~\cite{Mingwu:PRL:2011} or
Erbium~\cite{Aikawa:PRL:2012}. However, in these systems, the regime in which dipolar interactions dominate is restricted to low values
of the gas parameter, $\rho a^3 \lesssim 10^{-4}$ (with $\rho$ the density and $a$ the s-wave scattering length), where the system is very
dilute.
Increasing $n$ leads to the enhancement of three-body losses, while increasing 
$a$ implies that short-range interactions take over the DDI, thus washing out 
dipolar effects.

The physics of dipolar quantum gases can be drastically changed if one 
considers dipolar molecules instead of 
magnetic atoms. This is because dipolar molecules have a much higher dipolar 
strength (from $10$ to $1000$ times bigger than in $^{164}$Dy), thus allowing for a drastic increase of $a$ (and consequently, the gas parameter) without
losing dipolar physics. The experimental achievement of a BEC of
molecules~\cite{Bigagli2024} also represents an excellent opportunity 
to benchmark the validity of Bogoliubov theory and the finite-temperature 
extended Gross Pitaevskii equation (TeGPE) for the theoretical description of 
these novel systems. Remarkably, recent results in the strongly correlated, 
dipole-dominated limit show the failure of
the eGPE to reproduce quantum Monte Carlo results for the critical number required to form molecular droplets~\cite{langen_prl}. However, it has
not been addressed how accurate the TeGPE is in the specific conditions of the 
experiment of Ref.~\cite{Bigagli2024}, which lays far away from the highly
correlated regime. 

Temperature is a key parameter in
the theoretical description of the molecular condensate achieved in the experiment, since only a maximum condensate
fraction of $~\sim 60 \%$ was achieved. Therefore, there is an
important thermal depletion of the BEC that the state-of-the-art theoretical 
approach, the zero temperature extended Gross-Pitaevskii equation, is not able 
to describe~\cite{Giorgini:JLTP:1997,boudjemaa:aop:2017, boudjemaa:pra:2018, oktel19,oktel20,baena22}. In fact, recent experimental and theoretical results have shown
that thermal effects play a rather counterintuitive role in dipolar systems
by the promotion of supersolidity by heating~\cite{Ferlaino:PRL:2021,baena22,baena24,He2024}.
Nevertheless, the study of dipolar systems at finite temperature is a rather
unexplored topic, and as a consequence it is important to test and benchmark the 
regime of validity of the available theoretical tools, especially in the 
context of polar molecules.

\begin{figure*}[t]
\centering
\includegraphics[width=\linewidth]{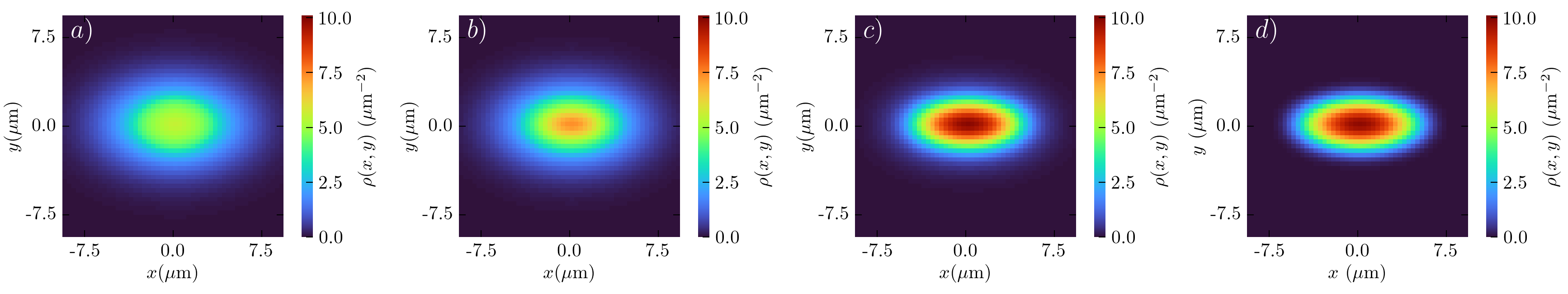}
\caption{
Emergence of a molecular Bose Einstein Condensate as the temperature is lowered.
The panels depict the column density $\rho(x,y) = \int dz  \left( \abs{\psi({\bf 
r})}^2 + \rho_d({\bf r}) \right)$, with $\psi({\bf r})$ the condensate wave function
and $\rho_d({\bf r})$ the density of depleted molecules. The total number of molecules is
$N=298$, while the temperatures are $T = 10.5, 9.5, 6$ and $0$ nK for panels
$a)$ to $d)$, respectively. The condensate fractions for each case are $f_c 
=0.01, 0.12, 0.58$ and $0.96$, respectively. Panel $c)$ corresponds to the 
coldest point of the experiment of Ref.~\cite{Bigagli2024}.
}
\label{fig0}
\end{figure*}

\section{\label{sec:system}System and methods}

\subsection{The temperature extended Gross Pitaevskii equation (TeGPE)}

In this work, we employ Bogoliubov theory and the local density approximation (LDA) to account for  thermal fluctuations on the condensate~\cite{oktel19,oktel20,baena22}. We aim to study ensembles of NaCs molecules under the same conditions of the recent
experiment of Ref.~\cite{Bigagli2024}.
We employ the temperature dependent extended Gross-Pitaevskii equation
(TeGPE) for the condensate wave function $\psi({\bf r})$
given by
\begin{eqnarray}
 \mu \psi({\bf r}) & = & \bigg(-\frac{\hbar^2\nabla^2}{2m} + U({\bf r}) + 
\!\!\int \!{\rm d}{\bf r}^\prime V({\bf r}-{\bf r}') \label{TeGPE} \\
  & \times  & \left[ \abs{\psi({\bf 
r}')}^2 + \rho_d({\bf r}') \right] + H_{\rm qu}({\bf r}) + H_{\rm th}({\bf r})
\bigg) \psi({\bf r}) \nonumber \ .
\end{eqnarray}
In this expression, $\mu$ is the chemical potential, $m$ is the molecular mass, 
$V({\bf r}) = V_{\rm dd}({\bf r}) + \frac{4\pi\hbar^2 a}{m} \delta ({\bf r})$ 
stands for both the contact and dipole-dipole interactions,
$a$ is the s-wave scattering
length, and $\rho_d({\bf r})$ is the density of molecules out of the BEC state. The term $U({\bf r}) = \frac{1}{2} m \left( \omega_x^2 x^2 + \omega_y^2 y^2 + \omega_z^2 z^2 \right)$ is the harmonic trap. In the experiment of Ref.~\cite{Bigagli2024}, the orientation of the molecular dipoles could not be determined. Nevertheless, we consider the dipoles to be polarized along the $z$-axis, and show that our results are robust with respect to variations of this direction (see Sec.~\ref{sec:interaction}).
The dipole-dipole interaction is given by
\begin{align}
 V_{\rm dd}({\bf r}) = \frac{d^2_{\rm eff}}{4 \pi \epsilon_0} \left( \frac{1 - 3 \cos^2 \theta}{r^3}  \right) \ ,
 \label{eq_ddi}
\end{align}
where ${\bf r} = {\bf r}_i - {\bf r}_j$, $r = \abs{{\bf r}}$, $\theta$ is the
angle between the vector ${\bf r} $ and the $z$ axis, $d_{\rm eff}$ is the effective dipole moment and $\epsilon_0$ is the permittivity of free space.
The
terms $H_{\rm qu}$ and $H_{\rm th}$ of Eq.~\ref{TeGPE} account for the effect of quantum and
thermal fluctuations, respectively. The first term is given by the usual 
Lee-Huang-Yang expression~\cite{Lima:2011eq,pelster12}
\begin{equation}
 H_{\rm qu}({\bf r}) = \frac{32}{3 \sqrt{\pi}} g \sqrt{a^3} Q_5(a_{\rm dd}/a) 
\abs{ \psi({\bf r}) }^3 \ .
\end{equation}
The parameter $a_{\rm dd} = m d^2_{\rm eff}/(12\pi \hbar^2 \epsilon_0)$ corresponds to the dipole length and the
auxiliary function $Q_5(a_{\rm dd}/a)$ is given by~\cite{pelster12}
\begin{equation}
 Q_5(a_{\rm dd}/a_s) = \int_{0}^1 du \left( 1 - \frac{a_{\rm dd}}{a_s} + 3 
\left( \frac{a_{\rm dd}}{a_s} \right) u^2 \right)^{5/2} \ .
\end{equation}
The correction by thermal fluctuations in Eq.~\ref{TeGPE} is given by
\begin{align}
 H_{\rm th}({\bf r}) &= { \int \frac{d{\bf k}}{(2 \pi)^3} \frac{ V({\bf k}) }{ 
\left( e^{\beta \varepsilon_{{\bf k}}}-1 \right) } \left[ \abs{u_{\bf k}}^2 + 
\abs{v_{\bf k}}^2 - 2 \abs{u_{\bf k}} \abs{v_{\bf k}}  \right] } \ , \label{TF}
\end{align}
where $V({\bf k})$ is the Fourier transform of the DDI plus contact potential, $u_{\bf k}$ and $v_{\bf k}$ are the Bogoliubov amplitudes while $\beta = 1/k_B T$.

\subsection{Calculation of the Bogoliubov amplitudes}

We take the following considerations in the calculation of the thermal correction to the eGPE, i.e. the $H_{\rm th}({\bf r})$ term of Eq.~\ref{TF}. The fact that the experimental condensate that we aim to describe lies in the regime $a>a_{\rm dd}$, meaning that the system is stable at the mean field level, allows us to account for the curvature of the condensate for vanishing densities, i.e. near the edges of the trap. This allows us to modify the usual calculation under the LDA for dipolar systems~\cite{oktel19,oktel20,baena22} to obtain a convergent thermal depletion.
Under these circumstances, the Bogoliubov de-Gennes equations for the excitations are
\begin{align}
 E_{\bf k} u_{\bf k}({\bf r}) &= \left( \frac{\hbar^2 k^2}{2 m} + U({\bf r}) - \mu_{\rm MF} \right) u_{\bf k}({\bf r}) \nonumber \\
 &+ u_{\bf k}({\bf r}) \!\!\int \!{\rm d}{\bf r}^\prime V({\bf r}-{\bf r}') \abs{\psi_{\rm MF}({\bf r}')}^2 \nonumber \\
 &+ \abs{\psi_{\rm MF}({\bf r})}^2 u_{\bf k}({\bf r}) V( {\bf k} ) - \abs{\psi_{\rm MF}({\bf r})}^2 v_{\bf k}({\bf r}) V( {\bf k} )
  \\
 - E_{\bf k} v_{\bf k}({\bf r}) &= \left( \frac{\hbar^2 k^2}{2 m} + U({\bf r}) - \mu_{\rm MF} \right) v_{\bf k}({\bf r}) \nonumber \\
 &+ v_{\bf k}({\bf r}) \!\!\int \!{\rm d}{\bf r}^\prime V({\bf r}-{\bf r}') \abs{\psi_{\rm MF}({\bf r}')}^2 \nonumber \\
 &+ \abs{\psi_{\rm MF}({\bf r})}^2 v_{\bf k}({\bf r}) V( {\bf k} ) - \abs{\psi_{\rm MF}({\bf r})}^2 u_{\bf k}({\bf r}) V( {\bf k} )  \ .
\end{align}
In this expression, $\mu_{\rm MF}$ is the mean field chemical potential,
and $\abs{\psi_{\rm MF}({\bf r})}^2$ is the mean-field density, i.e. the solution of Eq.~\ref{TeGPE} without accounting for beyond mean-field fluctuations. We can now use the exact identity for the mean field chemical potential
\begin{align}
 \mu_{\rm MF} &= -\frac{\hbar^2 }{2 m} \frac{\nabla^2 \psi_{\rm MF}}{\psi_{\rm MF}({\bf r})} + U({\bf r}) + \!\!\int \!{\rm d}{\bf r}^\prime V({\bf r}-{\bf r}') \abs{\psi_{\rm MF}({\bf r}')}^2
\end{align}
to reach the equations
\begin{align}
 &E_{\bf k} u_{\bf k}({\bf r}) = \left( \frac{\hbar^2 k^2}{2 m} + \frac{\hbar^2 }{2 m} \frac{\nabla^2 \psi_{\rm MF}}{\psi_{\rm MF}({\bf r})} \right) u_{\bf k}({\bf r}) \nonumber \\
 &+ \abs{\psi_{\rm MF}({\bf r})}^2 u_{\bf k}({\bf r}) V( {\bf k} ) - \abs{\psi_{\rm MF}({\bf r})}^2 v_{\bf k}({\bf r}) V( {\bf k} )
\label{eq_bdg_1} \\
 &- E_{\bf k} v_{\bf k}({\bf r}) = \left( \frac{\hbar^2 k^2}{2 m} + \frac{\hbar^2 }{2 m} \frac{\nabla^2 \psi_{\rm MF}}{\psi_{\rm MF}({\bf r})} \right) v_{\bf k}({\bf r}) \nonumber \\
 &+ \abs{\psi_{\rm MF}({\bf r})}^2 v_{\bf k}({\bf r}) V( {\bf k} ) - \abs{\psi_{\rm MF}({\bf r})}^2 u_{\bf k}({\bf r}) V( {\bf k} ) \ \label{eq_bdg_2} ,
\end{align}
where $V( {\bf k} )$ denotes the Fourier transform of the interaction, i.e.
\begin{equation}
V( {\bf k} ) = \frac{4 \pi \hbar^2 a}{m} + \frac{4 \pi \hbar^2 a_{\text{dd}}}{m} \left( 3 \frac{k_z^2}{k^2} -1\right) \ .
\label{fourier_int}
\end{equation}
We now replace $\abs{\psi_{\rm MF}({\bf r})}^2 \rightarrow \rho_0$ and build a density dependent function $F(\rho_0) = \frac{\hbar^2 }{2 m} \frac{\nabla^2 \psi_{\rm MF}}{\psi_{\rm MF}({\bf r})}$ by expressing this quantity in terms of the mean-field density $\abs{\psi_{\rm MF}({\bf r})}^2$. Since the thermal correction ignoring the condensate curvature has been proven successful in the reproduction of experimental results on dipolar systems at finite temperature~\cite{baena22}, we assume a completely flat condensate near the center and only account for the curvature near the edges of the trap. This can be achieved by retaining the contribution of $F(\rho_0)$ for densities for which $F(\rho_0) > 0$ only. It turns out that in this regime $F(\rho_0)$ can be approximately fitted to the density functional form
\begin{equation}
 F(\rho_0) = c_1 \ln(\rho_0) + c_2
\end{equation}
with $c_1$ and $c_2$ fitting parameters. This expression is exact for the ground state wave function of an isotropic harmonic oscillator. Thus, we solve the coupled equations
\begin{align}
 &E_{\bf k} u_{\bf k} = \left( \frac{\hbar^2 k^2}{2 m} + \tilde{F}(\rho_0) \right) u_{\bf k} + \rho_0 u_{\bf k} V( {\bf k} ) - \rho_0 v_{\bf k} V( {\bf k} )
 \\
 &- E_{\bf k} v_{\bf k} = \left( \frac{\hbar^2 k^2}{2 m} + \tilde{F}(\rho_0) \right) v_{\bf k} + \rho_0 v_{\bf k} V( {\bf k} ) - \rho_0 u_{\bf k} V( {\bf k} ) \ .
\end{align}
where $\tilde{F}(\rho_0) = \text{Max.}\{ F(\rho_0) \text{, }0 \}$.
Using $u_{\bf k}$ and $v_{\bf k}$ we can compute the thermal correction of Eq.~\ref{TF}, where the ${\bf r}$ dependence stems from the substitution $\rho_0 \rightarrow \abs{\psi({\bf r})}^2$ in the excitation spectrum $\varepsilon_{{\bf k}} (\rho_0)$ and the Bogoliubov amplitudes.

\subsection{Calculation of the density of thermal molecules}

In order to solve Eq.~\ref{TeGPE} iteratively, we need to obtain the density of thermal molecules at each iteration. Within Bogoliubov theory, the thermal density of depleted molecules for an homogeneous system can be computed as (see Eq. 35 of Ref.~\cite{oktel19})
\begin{align}
 \rho_{\rm th} = \int \frac{ d{\bf k} }{(2 \pi)^3} \frac{1}{e^{\beta \epsilon_{\bf k}} - 1} \left( \abs{u_{\bf k}}^2 + \abs{v_{\bf k}}^2 \right) \ . \label{thermal_density}
\end{align}
In order to obtain the depleted cloud of thermal molecules from Eq.~\ref{TeGPE}, we compute $\rho_{\rm th}$ for different values of the condensate density of the homogeneous system, $\rho_0$, thus building a density functional. Then, we solve Eq.~\ref{TeGPE} iteratively through imaginary time propagation and evaluate the thermal density at each point in space by computing $\rho_{th} \left( \rho_0 = \abs{\psi({\bf r})}^2 \right)$ i.e. we replace the homogeneous density by the value of the trapped density, as it is typically done in the LDA..

\section{Results}

To match the conditions of the experiment of Ref.~\cite{Bigagli2024}, the dipolar length is $a_{\rm dd} = 1250 a_0$ while the $s$-wave scattering length is set to $a = 1500 a_0$, meaning that $\epsilon_{\rm dd} =
a_{\rm dd}/a = 0.83 < 1$, so that dipolar effects are not dominant. The trap frequencies are $(\omega_x,\omega_y,\omega_z) =
2\pi (23, 49, 58)$ Hz. In what follows, we employ the characteristic length and
energy scales set by $r_0 = 12 \pi a_{\rm dd}$ and $\epsilon = \hbar^2/(m
r_0^2)$.

\subsection{Condensate fraction}

By numerically solving the TeGPE we can obtain the condensate wave 
function $\psi({\bf r})$ as well as the depleted density $n_{\rm
d}({\bf r})$. Since the gas lies in the dilute regime ($\rho a^3 < \rho_{\rm peak} a^3
\simeq 7.2 \times 10^{-4}$ with $\rho_{\rm peak}$ the peak density), the quantum
depletion of the condensate due to interaction effects is significantly smaller than the thermal one, within the range of
experimental temperatures. 
With this theoretical approach, we can describe the 
formation of the dipolar molecular BEC as the temperature decreases, as shown in
Fig.~\ref{fig0}. The figure shows both the rise of a density peak
and the tightening of the density distribution as more and more molecules fall into 
the condensate.

\begin{figure}[t]
\centering
\includegraphics[width=0.95\linewidth]{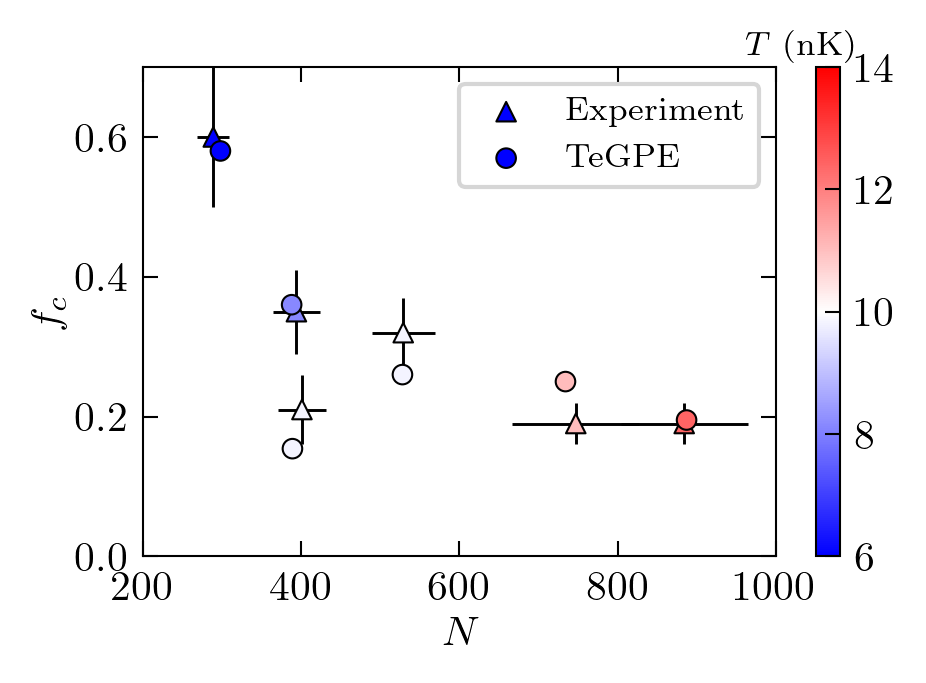}
\caption{
Condensate fraction of the NaCs molecular condensate under the experimental 
conditions of Ref.~\cite{Bigagli2024}. The horizontal axis corresponds to the 
total number of molecules while the colorbar indicates the temperature.
}
\label{fig_1}
\end{figure}

By fixing the total number of harmonically trapped molecules to
the values of the experiment of Ref.~\cite{Bigagli2024}, 
we can directly compare the TeGPE results for the condensate depletion with 
that measured in the experiment. We show the results in Fig.~\ref{fig_1}.
Notice that each point in the figure corresponds to a different total number of 
particles as well as a different temperature, since the experimental
measurements were taken during an evaporative cooling sequence. Remarkably, and 
as we can see from the figure, our result is in excellent agreement with
the experimental prediction within the errorbars, even when the condensate 
fraction is as low as $20 \%$.
The agreement is expected to worsen at higher temperatures, as the theory is expected to be more accurate at low temperatures, when less excited states
are populated.
The good agreement between the eGPE and the experimental observations indicates that the system is in the
universal regime, where the specific details of the molecular interaction do not significantly alter its properties. This means that, in the weakly dipolar regime, the parameter that controls the physics of the system is the scattering length $a$.
Furthermore, our results also stress the robustness of the finite temperature Bogoliubov theory for the
description of dipolar systems in the contact-dominated regime, where $a >
a_{\rm dd}$. This accuracy is still present in the dipolar dominated regime ($a 
< a_{\rm dd}$), where supersolids of $^{164}$Dy atoms arise~\cite{baena22}. 
However, further decreasing the scattering length in a system of dipolar 
molecules, which in turn leads to higher densities, breaks the validity of the
theory~\cite{langen_prl}.

It must be stressed that, while our condensed fractions are computed from the lowest free energy solution of the trapped system, those of the experiment of Ref.~\cite{Bigagli2024} are obtained after a 17 ms TOF expansion. However, due to the comparatively large lifetime of the condensate, of the order of seconds (see Fig. 5 of Ref.~\cite{Bigagli2024}), the condensate fraction should remain unchanged after this expansion, which allows us to compute it and find excellent agreement without the need to perform a TOF expansion.

\begin{table*}
    \begin{center}
        \begin{tabular}{ | c | c | c | c | }
        \hline
            ($N$, $T$) & $f_c^{\rm TeGPE}$ & $f_{c,\text{fit}}^{\rm TeGPE}$ & $f_{c}^{\rm exp.}$ \\ \hline
            ($289$, $6$ nK) & $0.58$ & $0.54$ & $0.6 \pm 0.1$ \\ \hline
            ($394$, $8.2$ nK) & $0.36$ & $0.28$ & $0.35 \pm 0.06$ \\ \hline
            ($884$, $12.4$ nK) & $0.195$ & $0.156$ & $0.19 \pm 0.03$ \\ \hline
        \end{tabular}
    \end{center}
    \caption{Condensate fraction obtained from the TeGPE calculations for 
    different values of $N$ and $T$ corresponding to the experiment of Ref.~\cite{Bigagli2024}. $f_c$
indicates the value computed by solving the TeGPE, $f_{c,\text{fit}}$ 
corresponds to the estimation applying the bimodal fit of Eq.~\ref{eq_bimodal} 
to the density profiles, and $f_{c}^{\rm exp.}$ denotes the
experimental values.}
    \label{table_1}
\end{table*}

In Ref.~\cite{Bigagli2024}, the condensate fraction is estimated by 
fitting a bimodal density distribution to the experimentally measured density 
profile after the expansion. The bimodal distribution is given by
\begin{align}
 \rho_{\rm fit}(x) &= \frac{15 N_{\rm TF}}{16 \sigma_{\rm TF}} \text{Max.}\left\{ 1 - \left( \frac{x - x_{\rm TF}}{\sigma_{\rm TF}} \right)^2 \text{, }0 \right\}^2 \nonumber \\
 &+ \frac{N_{\rm G}}{\sqrt{2 \pi} \sigma_{\rm G}} \exp\left( -\frac{(x - x_{\rm G})^2}{2 \sigma_{\rm G}^2} \right) \ ,
 \label{eq_bimodal}
\end{align}
where $N_{\rm TF}$, $N_{\rm G}$, $\sigma_{\rm TF}$, $\sigma_{\rm G}$, $x_{\rm 
TF}$ and $x_{\rm G}$ 
are fitting parameters. The first term on the r.h.s of Eq.~\ref{eq_bimodal} corresponds to the contribution of the condensate to the density while the second term accounts for the thermal density. Moreover, $\sigma_{\rm TF, G}$ represents the width of the
fit, $x_{\rm TF, G}$ the centers, and  $N_{\rm TF, G}$ stand for the number of condensed and thermally excited particles,
respectively. The condensate fraction is then estimated as $f_c \simeq N_{\rm 
TF}/(N_{\rm TF} + N_{\rm G})$. It is interesting to benchmark this procedure to obtain $f_c$ by
applying it to our theoretical density profiles and comparing the results with
our TeGPE estimation for the same quantity. We show the results in Table~\ref{table_1}.

\begin{figure}[t]
\centering
\includegraphics[width=0.8\linewidth]{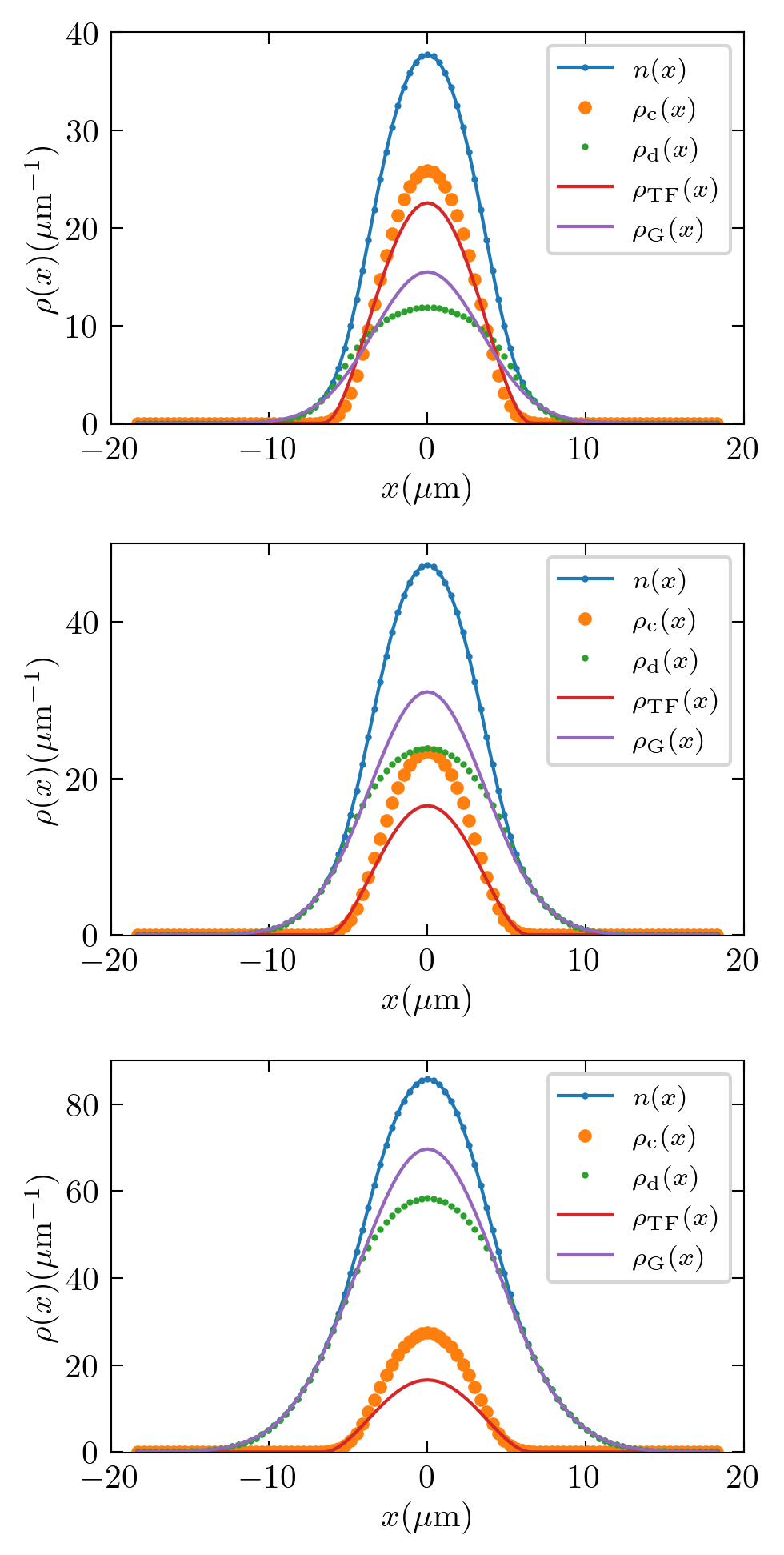}
\caption{
Ground state column density $\rho(x)$, condensate density $\rho_{\rm c}(x)$, and depleted density $\rho_{\rm d}(x)$ of the molecular condensate at three different points of the experiment of Ref.~\cite{Bigagli2024}: ($N=289$,$T=6$ nK) (top), ($N=394$,$T=8.2$ nK) (middle), ($N=884$,$T=12.6$ nK) (bottom). We also include the gaussian and TF contributions of the bimodal fit of Eq.~\ref{eq_bimodal}, which are obtained by fitting the total density.}
\label{fig_2}
\end{figure}

\begin{figure}[t]
\centering
\includegraphics[width=0.95\linewidth]{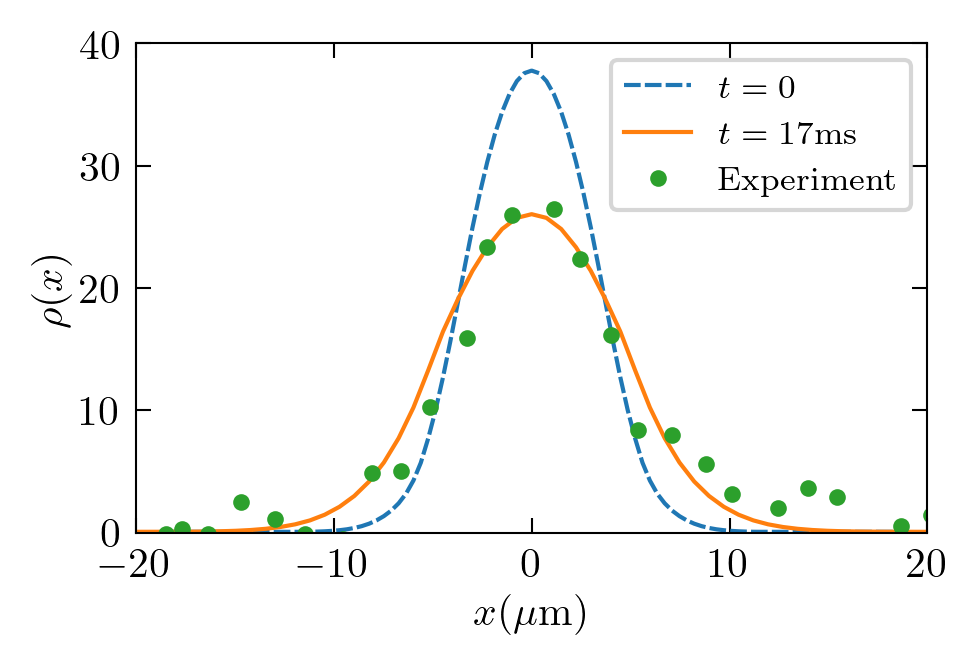}
\caption{ Ground state (dashed line) and time-evolved density profiles (solid
line) obtained, respectively, from the solution of the TeGPE and the tTeGPE. The time evolution corresponds to a 17 ms TOF expansion. Dots
indicate the experimental density distribution of Ref.~\cite{Bigagli2024}. The
number of particles is $N=289$ while the temperature is $T=6$ nK.}
\label{fig_5}
\end{figure}

By solving the TeGPE, we can obtain the ground state density profiles, together with the condensate and excited density distributions. We can use our results to further benchmark the bimodal fit of Eq.~\ref{eq_bimodal} by comparing the shape of the condensate and thermal components to the fitting functions. We denote the gaussian and TF components of the fit of Eq.~\ref{eq_bimodal} as $\rho_{\rm TF}(x)$ and $\rho_{\rm G}(x)$, respectively, and show our results in Fig.~\ref{fig_2}. From both the figure and the data of Table~\ref{table_1}, we see that the bimodal fit consistently and slightly underestimates the condensate fraction, although it lays indeed close
to the computed condensate fractions when applied to the TeGPE profiles.. Regardless, we see that the fit quantitatively captures the width of the condensate and depleted distributions, and also gives a solid approximation of the density at the center of each distribution.

Our formalism is capable of estimating the critical temperature of the condensate for a fixed total number of molecules $N$. This is done by solving Eq.~\ref{TeGPE} at different temperatures to find the one for which $N = N_{d}$, with $N_{d}$ the number of depleted molecules. To further characterize the coldest condensate obtained in the experiment of Ref.~\cite{Bigagli2024}, we perform calculations for a system of $N=289$ particles and find that the condensate should disappear at $T \simeq 9.8$ nK.

\subsection{Density profiles after TOF}

To further characterize the system, one can have a look at the molecular density profile.
As stated previously, in the experiment of Ref.~\cite{Bigagli2024}, the column densities of the molecular cloud are
measured
after a TOF expansion of 17 ms.  To compare our density profiles with the experimental ones, therefore, we need to simulate this TOF expansion. Solving
the TeGPE in the presence of the trap allows us to find the lowest free-energy 
quantum state. Subsequently, taking its solution as an initial state, we can
approximately simulate the TOF expansion of the system by solving the 
time-dependent TeGPE (tTeGPE) when the trap is removed, which is given by
\begin{align}
 i \hbar \pdv{\psi}{t} =& \left(-\frac{\hbar^2\nabla^2}{2m} + U({\bf r})  \right. \nonumber \\ +
 &\left. \!\!\int \!{\rm d}{\bf r}^\prime V({\bf r}-{\bf r}') \left( \abs{\psi({\bf r}',t)}^2 + \rho_d({\bf r}',t) \right) \right. \nonumber \\
 &\left. + H_{\rm qu}({\bf r},t) + H_{\rm th}({\bf r},t) \right) \psi({\bf r},t) \label{tTeGPE} \\
 &= \hat{K} \psi({\bf r},t) + \hat{V}_{\rm TeGPE}({\bf r},t) \psi({\bf r},t) \ .
\end{align}
where $\hat{K} = \frac{\hat{P}^2}{2 m}$ is the kinetic energy operator. In principle, this equation must be solved simultaneously while numerically performing the time evolution of the non-condensed density $\rho_d({\bf r},t)$, which is governed by the time-dependent Bogoliubov equations. However, the depleted cloud is obtained under the local density approximation, which ignores the trapping potential away from the low-density regions. Since the Hamiltonian that is used in the time-evolution has the trapping potential removed, we assume that the depleted cloud remains unchanged during the TOF expansion, and compute the time evolution of the condensate component by solving Eq.~\ref{tTeGPE} for the static distribution of non-condensed molecules.

In practice, to implement the real-time evolution, we apply the time-evolution operator to the wave function iteratively, which we split as
\begin{align}
 \hat{O}_t \simeq \exp\left[ -i \frac{ \hat{V}_{\rm TeGPE} \Delta t }{ 2 \hbar} \right] \exp\left[ -i \frac{ \hat{K} \Delta t }{ \hbar} \right] \exp\left[ -i \frac{ \hat{V}_{\rm TeGPE} \Delta t }{ 2 \hbar} \right] \ ,
\end{align}
which is exact up to order $\Delta t^2$. If the time step is small enough, we can use the initial state at every iteration to compute $V_{\rm TeGPE}({\bf r},t)$, which depends on the density. The terms of the propagator which contain the potential are applied in position space, while the kinetic term is applied in momentum space by using an FFT algorithm.

We show in
Fig.~\ref{fig_5} the tTeGPE result after the TOF expansion, together with the TeGPE
ground state and post-TOF experimental density distributions, for the
coldest point recorded in the experiment ($N=289$, $T=6$ nK). Remarkably, we
obtain an excellent agreement between theory and experiment. Moreover, we see that, for the case shown, where 
the condensate fraction is $f_c \simeq 0.6$, the expansion dynamics are mostly 
governed by the molecules in the condensate, despite the presence of a significant
thermal cloud.

\subsection{Peak density}

\begin{figure}[t]
\centering
\includegraphics[width=0.95\linewidth]{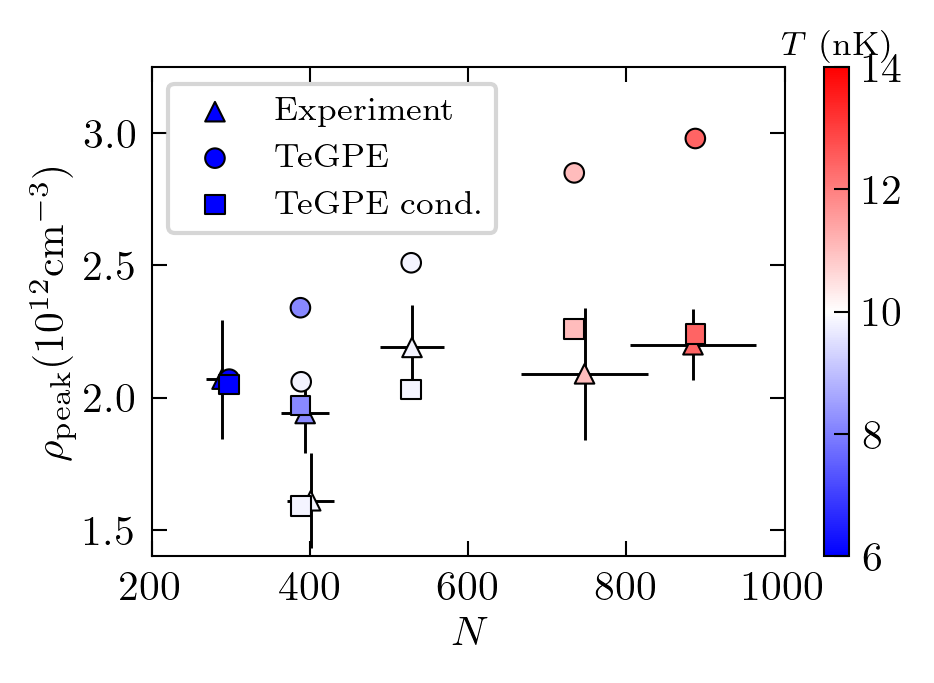}
\caption{
Peak density of the molecular condensate under the experimental conditions
of Ref.~\cite{Bigagli2024}. The horizontal axis corresponds to the total number
of particles while the colorbar indicates the temperature. The ``TeGPE'' results correspond to the peak density of the total cloud obtained from a TeGPE simulation, while the ``TeGPE cond.'' results correspond to the peak density of the condensate molecules alone. Note that the
experimental estimation does not account for the effect of the DDI nor the thermal cloud~\cite{Bigagli2024,dalfovo99}.}
\label{fig_3}
\end{figure}

 We estimate the peak density from 
the ground state density profiles (no TOF expansion is performed) and compare it to the experimental
results in Fig.~\ref{fig_3}. It must be remarked that the authors in Ref.~\cite{Bigagli2024} use the number of condensed molecules after the TOF expansion to obtain the peak density of the system before the expansion. They use the expression
\begin{align}
 \rho_{\rm peak} = \frac{15^{2/5}}{8 \pi} \left[ m^3 \overline{\omega}^3 N_{\rm c} / \left( \hbar^3 a^{3/2} \right) \right]^{2/5} \ .
 \label{peak_dens_exp}
\end{align}
where $N_c$ is the number of condensed molecules and $\overline{\omega} = \sqrt{\omega_x \omega_y \omega_z}$ is the geometric mean of the trapping frequencies. This expression ignores both the effect of the DDI, as well as the presence of the thermal cloud~\cite{dalfovo99}.
From the results, we see that the TeGPE estimates
lay relatively close to the experimental ones for the highest values of the condensate 
fraction, while the agreement worsens for the largest temperatures/lowest 
condensate fractions, mainly due to the experimental results neglecting the thermal cloud. Because of this, we also include the values for the peak density of the condensate alone (squares in Fig.~\ref{fig_3}). In this case, we find excellent agreement with the experimental results at all temperatures.

\subsection{\label{sec:interaction}The pseudopotential and the inter-molecular interaction under double MW shielding}

The full nature of the inter-molecular interaction in the experiment is not disclosed in Ref.~\cite{Bigagli2024}. However, Refs.~\cite{karman2025:arxiv,deng2025:arxiv} shed more light into this question. The long range part of the interaction between molecules under a circularly polarized MW field $\sigma^+$ and a linearly polarized MW field $\pi$ can be expanded in spherical harmonics (see Eq. 24 of Ref.~\cite{karman2025:arxiv})
\begin{align}
 V_{\rm long} ({\bf r}) = -2 \sum_{m} (d_{\rm eff}^m)^2 \frac{Y_2^m (\theta,\phi)}{4 \pi \epsilon_0 r^3} \label{interaction}
\end{align}
where ${\bf r}$ is the relative position vector between two molecules, $\theta$ and $\phi$ are the polar and azimuthal angles and $Y_2^m (\theta,\phi)$ is the $l=2, m$ spherical harmonic. The $z$-axis is taken here as the direction of the polarization. Using the convention of Ref.~\cite{karman2025:arxiv}, the effective dipole moment in Eq.~\ref{interaction} can be either a real or an imaginary number, depending on the sign of the effective dipole length $a_{\rm dd}^m = m (d^m_{\rm eff})^2/(12 \pi \hbar^2 \epsilon_0)$ associated to each partial wave channel. In the absence of ellipticity, the interaction is just given by the $m=0$ term, while the magnitude and sign of the effective dipole length $a_{\rm dd}^0$ can be changed by tuning the Rabi frequencies and detunings of the MW fields (it can also be completely cancelled)~\cite{karman2025:arxiv, deng2025:arxiv}. Thus, in this case, the long range part of the interaction corresponds to the standard dipole-dipole interaction (DDI) for dipoles polarized along the $z$-axis for $a_{\rm dd}^0 > 0$ and the anti-dipolar interaction for $a_{\rm dd}^0 < 0$. In this sense, the interaction employed in our work matches the inter-molecular interaction for zero ellipticity and $a_{\rm dd}^0 > 0$. In this case, we can relate the effective dipole moment $d_{\rm eff}^{(0)}$ with the Rabi frequencies and detunings employed in the experiments to achieve the double MW shielding. By comparing Eq.~\ref{interaction} with Eq. (21) of Ref.~\cite{deng2025:arxiv} we obtain
\begin{align}
 d_{\rm eff}^{(0)} = \sqrt{ -\frac{1}{3} \sqrt{\frac{\pi}{5}} d^2 (3 \cos (2 \beta)-1) \cos^2 \alpha \sin^2 \alpha } \ .
\end{align}
where $d$ is the bare dipole moment of the molecules and $\alpha$ and $\beta$ are two Euler angles, which depend on the detunings ($\delta_\sigma$, $\delta_\pi$) and Rabi frequencies ($\Omega_\sigma$, $\Omega_\pi$) of the $\sigma^+$ and $\pi$ polarized fields. These Euler angles arise from the diagonalization of the single molecule Hamiltonian under the presence of the aforementioned fields and their dependence on the parameters $\Omega_\sigma, \Omega_\pi, \delta_\sigma, \delta_\pi$ is not analytical.

\begin{figure}[t]
\centering
\includegraphics[width=0.9\linewidth]{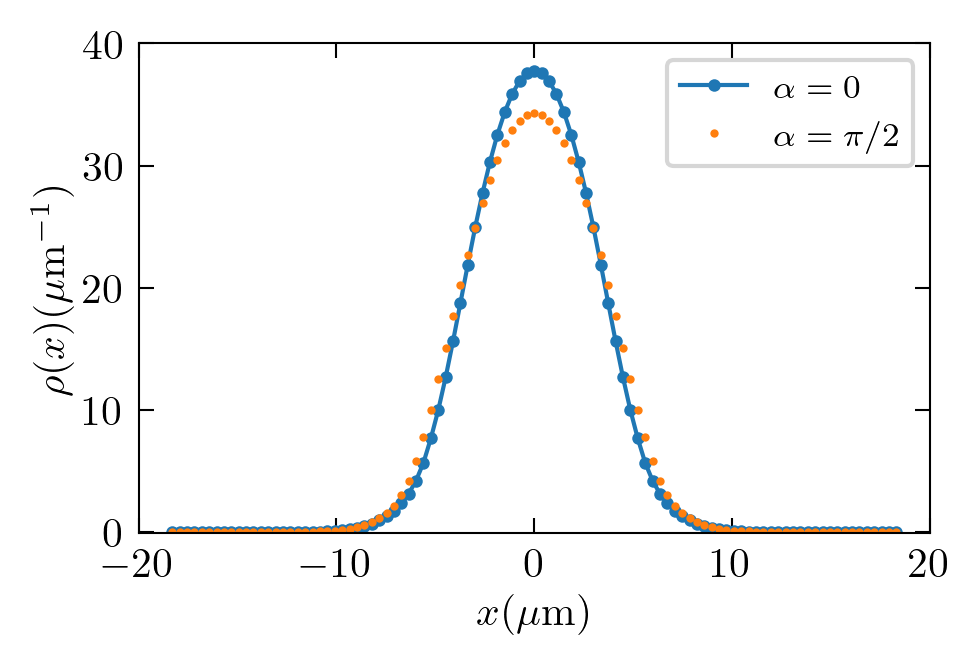}
\caption{
Ground state density distribution of the molecular condensate for the coldest experimental point ($N=289$, $T=6$ nK, $\epsilon_{\rm dd} = 0.83$) for a system of dipoles polarized along the $z$-axis ($\alpha = 0$, blue line with markers), and for a system polarized along the $x$-axis ($\alpha = \pi/2$, orange markers). Here $\alpha$ is the tilting angle of the dipoles in the $x$-$z$ plane}
\label{fig_4}
\end{figure}

While we have assumed the dipoles to be polarized along the $z$-axis, the polarization direction of the dipoles in the experiment is unknown, as mentioned in the supplementary information of Ref.~\cite{Bigagli2024}.
In order to quantify the effect of the polarization in our results, we have performed calculations at the coldest experimental point ($N=289$,$T=6$ nK)
for a system of dipoles polarized along the $x$-axis (see Fig.~\ref{fig_4} ). Our results show that the TeGPE density profile is only slightly affected by drastically changing the polarization. Moreover, the condensate depletion is only modified by $\simeq 0.01$.
This means that the uncertainty in the polarization direction does not change the physical picture presented in this work.

Lastly, the short range part of the full inter-molecular potential corresponds to a term decaying with the inter-molecular distance as $\sim 1/r^6$ and anisotropic, with a dependence on $\theta$ at zero ellipticity~\cite{deng2025:arxiv}. In our work, we have considered a pseudopotential of the form~\cite{langen_prl,boudjemaa:pra:2025}
\begin{align}
 V({\bf r}) = \frac{4 \pi \hbar^2 a}{m} \delta \left( {\bf r} \right) + V_{\rm dd} ({\bf r})
\end{align}
with $V_{\rm dd} ({\bf r})$ given in Eq.~\ref{eq_ddi}. This is a good approximation as long as the generalized scattering lengths, $a_l^{m=0}$ are insensitive to the anisotropy of the short range potential for $l > 0$ (see Ref.~\cite{Yi2001}).
While we can not explicitly check this, because we do not have access to the full inter-molecular potential, we performed T-matrix calculations employing the inter-molecular potential under a single, circularly polarized MW field of Ref.~\cite{Yi2001} and checked that these scattering lengths are unaffected by the short range anisotropy. Therefore, it is reasonable to assume the same will happen for the doubly MW shielded interaction at zero ellipticity, since the dependence on the inter-molecular distance of the short and long range parts of the interaction is the same, and there is only anisotropy through a dependence on $\theta$ in both cases. Thus, we can argue that the pseudopotential employed in our work is accurate enough to describe the experimental system.

\section{Conclusions}

In conclusion, we have employed the temperature-dependent extended 
Gross-Pitaevskii 
equation to theoretically model the condensate of dipolar molecules of Ref.~\cite{Bigagli2024}. We have reported calculations for the condensate
fraction that are in excellent agreement with the experimental values. We have
also obtained excellent agreement with the experimental density profile
after a 17 ms TOF expansion in the coldest experimental conditions, where the
condensate fraction is the highest.
Moreover, we have reported results for the peak density of the condensate component, and find excellent agreement between our results and the experimental ones, which neglect the presence of the thermal cloud. We also provide the values of the peak density accounting for the presence of such thermal density of depleted molecules.

Our results, thus, establish the validity of Bogoliubov theory and the TeGPE for 
the 
description of the very novel molecular condensates in the dilute regime, away 
from strong dipolar interactions, which need exact \textit{ab-initio}
methods like quantum Monte Carlo algorithms to properly be described
~\cite{langen_prl}. Even in the regime considered in this work,
slight quantitative improvements in regards to the density profiles and peak 
densities could be achieved by means of quantum Monte Carlo simulations, which 
are able to exactly account for the excited thermal cloud. Moreover, the use of 
these methods can also allow for the quantification of the superfluid density 
present in the system.

We believe that the control of dipolar molecular condensates will open interesting research avenues in the near future. Future work should address the study of dipolar molecules in the strongly dipolar regime, where the enhanced dipolar moment will play a dominant role. Although it is difficult to anticipate all the possibilities that these systems will offer, a direct comparison with magnetic atom setups can give us some insight. Firstly, as it happens with dipolar atoms, the formation of self-bound molecular clusters (droplets) is expected to occur. However, the critical number of molecules to form them can be up to ten times smaller than the one observed in experiments with Dy and Er atoms~\cite{langen_prl}. And secondly, while a supersolid state formed by arrays of atomic clusters has been experimentally realized, it is still unclear if the smaller, and more correlated, molecular dipolar clusters are able to maintain phase coherence between them or, on the contrary, they will form an insulating state. Finally, in dipolar condensates of magnetic atoms, it has been observed that a supersolid state can be reached after heating a gaseous BEC by increasing the temperature in the nK regime~\cite{Ferlaino:PRL:2021}.
Thus, a third important line of research will be to elucidate whether the balance between thermal fluctuations, quantum correlation and anisotropic interaction can lead to the observation of similar exotic phenomena in molecular dipolar condensate.

We acknowledge financial support from Ministerio de Ciencia e 
Innovaci\'on
MCIN/AEI/10.13039/501100011033
(Spain) under Grant No. PID2023-147469NB-C21  and
from AGAUR-Generalitat de Catalunya Grant No. 2021-SGR-01411.
R.B. acknowledges funding from ADAGIO (Advanced Manufacturing Research Fellowship Programme in the Basque – New Aquitaine Region)
MSCA COFUND Post-Doctoral fellowship programme (Grant agreement No. 101034379).

\bibliography{paper_bec_molecules}

\begin{thebibliography}{38}%
\makeatletter
\providecommand \@ifxundefined [1]{%
 \@ifx{#1\undefined}
}%
\providecommand \@ifnum [1]{%
 \ifnum #1\expandafter \@firstoftwo
 \else \expandafter \@secondoftwo
 \fi
}%
\providecommand \@ifx [1]{%
 \ifx #1\expandafter \@firstoftwo
 \else \expandafter \@secondoftwo
 \fi
}%
\providecommand \natexlab [1]{#1}%
\providecommand \enquote  [1]{``#1''}%
\providecommand \bibnamefont  [1]{#1}%
\providecommand \bibfnamefont [1]{#1}%
\providecommand \citenamefont [1]{#1}%
\providecommand \href@noop [0]{\@secondoftwo}%
\providecommand \href [0]{\begingroup \@sanitize@url \@href}%
\providecommand \@href[1]{\@@startlink{#1}\@@href}%
\providecommand \@@href[1]{\endgroup#1\@@endlink}%
\providecommand \@sanitize@url [0]{\catcode `\\12\catcode `\$12\catcode
  `\&12\catcode `\#12\catcode `\^12\catcode `\_12\catcode `\%12\relax}%
\providecommand \@@startlink[1]{}%
\providecommand \@@endlink[0]{}%
\providecommand \url  [0]{\begingroup\@sanitize@url \@url }%
\providecommand \@url [1]{\endgroup\@href {#1}{\urlprefix }}%
\providecommand \urlprefix  [0]{URL }%
\providecommand \Eprint [0]{\href }%
\providecommand \doibase [0]{http://dx.doi.org/}%
\providecommand \selectlanguage [0]{\@gobble}%
\providecommand \bibinfo  [0]{\@secondoftwo}%
\providecommand \bibfield  [0]{\@secondoftwo}%
\providecommand \translation [1]{[#1]}%
\providecommand \BibitemOpen [0]{}%
\providecommand \bibitemStop [0]{}%
\providecommand \bibitemNoStop [0]{.\EOS\space}%
\providecommand \EOS [0]{\spacefactor3000\relax}%
\providecommand \BibitemShut  [1]{\csname bibitem#1\endcsname}%
\let\auto@bib@innerbib\@empty
\bibitem [{\citenamefont {Gorshkov}\ \emph {et~al.}(2008)\citenamefont
  {Gorshkov}, \citenamefont {Rabl}, \citenamefont {Pupillo}, \citenamefont
  {Micheli}, \citenamefont {Zoller}, \citenamefont {Lukin},\ and\ \citenamefont
  {B\"uchler}}]{Gorshkov08}%
  \BibitemOpen
  \bibfield  {author} {\bibinfo {author} {\bibfnamefont {A.~V.}\ \bibnamefont
  {Gorshkov}}, \bibinfo {author} {\bibfnamefont {P.}~\bibnamefont {Rabl}},
  \bibinfo {author} {\bibfnamefont {G.}~\bibnamefont {Pupillo}}, \bibinfo
  {author} {\bibfnamefont {A.}~\bibnamefont {Micheli}}, \bibinfo {author}
  {\bibfnamefont {P.}~\bibnamefont {Zoller}}, \bibinfo {author} {\bibfnamefont
  {M.~D.}\ \bibnamefont {Lukin}}, \ and\ \bibinfo {author} {\bibfnamefont
  {H.~P.}\ \bibnamefont {B\"uchler}},\ }\bibfield  {title} {\enquote {\bibinfo
  {title} {Suppression of inelastic collisions between polar molecules with a
  repulsive shield},}\ }\href {\doibase 10.1103/PhysRevLett.101.073201}
  {\bibfield  {journal} {\bibinfo  {journal} {Phys. Rev. Lett.}\ }\textbf
  {\bibinfo {volume} {101}},\ \bibinfo {pages} {073201} (\bibinfo {year}
  {2008})}\BibitemShut {NoStop}%
\bibitem [{\citenamefont {Lin}\ \emph {et~al.}(2023)\citenamefont {Lin},
  \citenamefont {Chen}, \citenamefont {Jin}, \citenamefont {Shi}, \citenamefont
  {Deng}, \citenamefont {Zhang}, \citenamefont {Qu\'em\'ener}, \citenamefont
  {Shi}, \citenamefont {Yi},\ and\ \citenamefont {Wang}}]{Junyu2023}%
  \BibitemOpen
  \bibfield  {author} {\bibinfo {author} {\bibfnamefont {Junyu}\ \bibnamefont
  {Lin}}, \bibinfo {author} {\bibfnamefont {Guanghua}\ \bibnamefont {Chen}},
  \bibinfo {author} {\bibfnamefont {Mucan}\ \bibnamefont {Jin}}, \bibinfo
  {author} {\bibfnamefont {Zhaopeng}\ \bibnamefont {Shi}}, \bibinfo {author}
  {\bibfnamefont {Fulin}\ \bibnamefont {Deng}}, \bibinfo {author}
  {\bibfnamefont {Wenxian}\ \bibnamefont {Zhang}}, \bibinfo {author}
  {\bibfnamefont {Goulven}\ \bibnamefont {Qu\'em\'ener}}, \bibinfo {author}
  {\bibfnamefont {Tao}\ \bibnamefont {Shi}}, \bibinfo {author} {\bibfnamefont
  {Su}~\bibnamefont {Yi}}, \ and\ \bibinfo {author} {\bibfnamefont {Dajun}\
  \bibnamefont {Wang}},\ }\bibfield  {title} {\enquote {\bibinfo {title}
  {Microwave shielding of bosonic narb molecules},}\ }\href {\doibase
  10.1103/PhysRevX.13.031032} {\bibfield  {journal} {\bibinfo  {journal} {Phys.
  Rev. X}\ }\textbf {\bibinfo {volume} {13}},\ \bibinfo {pages} {031032}
  (\bibinfo {year} {2023})}\BibitemShut {NoStop}%
\bibitem [{\citenamefont {Karam}\ \emph {et~al.}(2023)\citenamefont {Karam},
  \citenamefont {Vexiau}, \citenamefont {Bouloufa-Maafa}, \citenamefont
  {Dulieu}, \citenamefont {Lepers}, \citenamefont {Borgloh}, \citenamefont
  {Ospelkaus},\ and\ \citenamefont {Karpa}}]{Karam2023}%
  \BibitemOpen
  \bibfield  {author} {\bibinfo {author} {\bibfnamefont {Charbel}\ \bibnamefont
  {Karam}}, \bibinfo {author} {\bibfnamefont {Romain}\ \bibnamefont {Vexiau}},
  \bibinfo {author} {\bibfnamefont {Nadia}\ \bibnamefont {Bouloufa-Maafa}},
  \bibinfo {author} {\bibfnamefont {Olivier}\ \bibnamefont {Dulieu}}, \bibinfo
  {author} {\bibfnamefont {Maxence}\ \bibnamefont {Lepers}}, \bibinfo {author}
  {\bibfnamefont {Mara Meyer zum~Alten}\ \bibnamefont {Borgloh}}, \bibinfo
  {author} {\bibfnamefont {Silke}\ \bibnamefont {Ospelkaus}}, \ and\ \bibinfo
  {author} {\bibfnamefont {Leon}\ \bibnamefont {Karpa}},\ }\bibfield  {title}
  {\enquote {\bibinfo {title} {Two-photon optical shielding of collisions
  between ultracold polar molecules},}\ }\href {\doibase
  10.1103/PhysRevResearch.5.033074} {\bibfield  {journal} {\bibinfo  {journal}
  {Phys. Rev. Res.}\ }\textbf {\bibinfo {volume} {5}},\ \bibinfo {pages}
  {033074} (\bibinfo {year} {2023})}\BibitemShut {NoStop}%
\bibitem [{\citenamefont {Bigagli}\ \emph {et~al.}(2023)\citenamefont
  {Bigagli}, \citenamefont {Warner}, \citenamefont {Yuan}, \citenamefont
  {Zhang}, \citenamefont {Stevenson}, \citenamefont {Karman},\ and\
  \citenamefont {Will}}]{Bigagli2023}%
  \BibitemOpen
  \bibfield  {author} {\bibinfo {author} {\bibfnamefont {Niccol{\`o}}\
  \bibnamefont {Bigagli}}, \bibinfo {author} {\bibfnamefont {Claire}\
  \bibnamefont {Warner}}, \bibinfo {author} {\bibfnamefont {Weijun}\
  \bibnamefont {Yuan}}, \bibinfo {author} {\bibfnamefont {Siwei}\ \bibnamefont
  {Zhang}}, \bibinfo {author} {\bibfnamefont {Ian}\ \bibnamefont {Stevenson}},
  \bibinfo {author} {\bibfnamefont {Tijs}\ \bibnamefont {Karman}}, \ and\
  \bibinfo {author} {\bibfnamefont {Sebastian}\ \bibnamefont {Will}},\
  }\bibfield  {title} {\enquote {\bibinfo {title} {Collisionally stable gas of
  bosonic dipolar ground-state molecules},}\ }\href {\doibase
  10.1038/s41567-023-02200-6} {\bibfield  {journal} {\bibinfo  {journal}
  {Nature Physics}\ }\textbf {\bibinfo {volume} {19}},\ \bibinfo {pages}
  {1579--1584} (\bibinfo {year} {2023})}\BibitemShut {NoStop}%
\bibitem [{\citenamefont {Mukherjee}\ and\ \citenamefont
  {Hutson}(2024)}]{Mukherjee2024}%
  \BibitemOpen
  \bibfield  {author} {\bibinfo {author} {\bibfnamefont {Bijit}\ \bibnamefont
  {Mukherjee}}\ and\ \bibinfo {author} {\bibfnamefont {Jeremy~M.}\ \bibnamefont
  {Hutson}},\ }\bibfield  {title} {\enquote {\bibinfo {title} {Controlling
  collisional loss and scattering lengths of ultracold dipolar molecules with
  static electric fields},}\ }\href {\doibase 10.1103/PhysRevResearch.6.013145}
  {\bibfield  {journal} {\bibinfo  {journal} {Phys. Rev. Res.}\ }\textbf
  {\bibinfo {volume} {6}},\ \bibinfo {pages} {013145} (\bibinfo {year}
  {2024})}\BibitemShut {NoStop}%
\bibitem [{\citenamefont {Bigagli}\ \emph {et~al.}(2024)\citenamefont
  {Bigagli}, \citenamefont {Yuan}, \citenamefont {Zhang}, \citenamefont
  {Bulatovic}, \citenamefont {Karman}, \citenamefont {Stevenson},\ and\
  \citenamefont {Will}}]{Bigagli2024}%
  \BibitemOpen
  \bibfield  {author} {\bibinfo {author} {\bibfnamefont {Niccol{\`o}}\
  \bibnamefont {Bigagli}}, \bibinfo {author} {\bibfnamefont {Weijun}\
  \bibnamefont {Yuan}}, \bibinfo {author} {\bibfnamefont {Siwei}\ \bibnamefont
  {Zhang}}, \bibinfo {author} {\bibfnamefont {Boris}\ \bibnamefont
  {Bulatovic}}, \bibinfo {author} {\bibfnamefont {Tijs}\ \bibnamefont
  {Karman}}, \bibinfo {author} {\bibfnamefont {Ian}\ \bibnamefont {Stevenson}},
  \ and\ \bibinfo {author} {\bibfnamefont {Sebastian}\ \bibnamefont {Will}},\
  }\bibfield  {title} {\enquote {\bibinfo {title} {Observation of
  bose--einstein condensation of dipolar molecules},}\ }\href {\doibase
  10.1038/s41586-024-07492-z} {\bibfield  {journal} {\bibinfo  {journal}
  {Nature}\ }\textbf {\bibinfo {volume} {631}},\ \bibinfo {pages} {289--293}
  (\bibinfo {year} {2024})}\BibitemShut {NoStop}%
\bibitem [{\citenamefont {Langen}\ \emph {et~al.}(2025)\citenamefont {Langen},
  \citenamefont {Boronat}, \citenamefont {S\'anchez-Baena}, \citenamefont
  {Bomb\'{\i}n}, \citenamefont {Karman},\ and\ \citenamefont
  {Mazzanti}}]{langen_prl}%
  \BibitemOpen
  \bibfield  {author} {\bibinfo {author} {\bibfnamefont {Tim}\ \bibnamefont
  {Langen}}, \bibinfo {author} {\bibfnamefont {Jordi}\ \bibnamefont {Boronat}},
  \bibinfo {author} {\bibfnamefont {Juan}\ \bibnamefont {S\'anchez-Baena}},
  \bibinfo {author} {\bibfnamefont {Ra\'ul}\ \bibnamefont {Bomb\'{\i}n}},
  \bibinfo {author} {\bibfnamefont {Tijs}\ \bibnamefont {Karman}}, \ and\
  \bibinfo {author} {\bibfnamefont {Ferran}\ \bibnamefont {Mazzanti}},\
  }\bibfield  {title} {\enquote {\bibinfo {title} {Dipolar droplets of strongly
  interacting molecules},}\ }\href {\doibase 10.1103/PhysRevLett.134.053001}
  {\bibfield  {journal} {\bibinfo  {journal} {Phys. Rev. Lett.}\ }\textbf
  {\bibinfo {volume} {134}},\ \bibinfo {pages} {053001} (\bibinfo {year}
  {2025})}\BibitemShut {NoStop}%
\bibitem [{\citenamefont {Kadau}\ \emph {et~al.}(2016)\citenamefont {Kadau},
  \citenamefont {Schmitt}, \citenamefont {Wenzel}, \citenamefont {Wink},
  \citenamefont {Maier}, \citenamefont {Ferrier-Barbut},\ and\ \citenamefont
  {Pfau}}]{Pfau:nature:2016}%
  \BibitemOpen
  \bibfield  {author} {\bibinfo {author} {\bibfnamefont {Holger}\ \bibnamefont
  {Kadau}}, \bibinfo {author} {\bibfnamefont {Matthias}\ \bibnamefont
  {Schmitt}}, \bibinfo {author} {\bibfnamefont {Matthias}\ \bibnamefont
  {Wenzel}}, \bibinfo {author} {\bibfnamefont {Clarissa}\ \bibnamefont {Wink}},
  \bibinfo {author} {\bibfnamefont {Thomas}\ \bibnamefont {Maier}}, \bibinfo
  {author} {\bibfnamefont {Igor}\ \bibnamefont {Ferrier-Barbut}}, \ and\
  \bibinfo {author} {\bibfnamefont {Tilman}\ \bibnamefont {Pfau}},\ }\bibfield
  {title} {\enquote {\bibinfo {title} {Observing the rosensweig instability of
  a quantum ferrofluid},}\ }\href@noop {} {\bibfield  {journal} {\bibinfo
  {journal} {Nature}\ }\textbf {\bibinfo {volume} {530}},\ \bibinfo {pages}
  {194--197} (\bibinfo {year} {2016})}\BibitemShut {NoStop}%
\bibitem [{\citenamefont {Schmitt}\ \emph {et~al.}(2016)\citenamefont
  {Schmitt}, \citenamefont {Wenzel}, \citenamefont {B{\"o}ttcher},
  \citenamefont {Ferrier-Barbut},\ and\ \citenamefont
  {Pfau}}]{Pfau:nature2:2016}%
  \BibitemOpen
  \bibfield  {author} {\bibinfo {author} {\bibfnamefont {Matthias}\
  \bibnamefont {Schmitt}}, \bibinfo {author} {\bibfnamefont {Matthias}\
  \bibnamefont {Wenzel}}, \bibinfo {author} {\bibfnamefont {Fabian}\
  \bibnamefont {B{\"o}ttcher}}, \bibinfo {author} {\bibfnamefont {Igor}\
  \bibnamefont {Ferrier-Barbut}}, \ and\ \bibinfo {author} {\bibfnamefont
  {Tilman}\ \bibnamefont {Pfau}},\ }\bibfield  {title} {\enquote {\bibinfo
  {title} {Self-bound droplets of a dilute magnetic quantum liquid},}\
  }\href@noop {} {\bibfield  {journal} {\bibinfo  {journal} {Nature}\ }\textbf
  {\bibinfo {volume} {539}},\ \bibinfo {pages} {259--262} (\bibinfo {year}
  {2016})}\BibitemShut {NoStop}%
\bibitem [{\citenamefont {Ferrier-Barbut}\ \emph {et~al.}(2016)\citenamefont
  {Ferrier-Barbut}, \citenamefont {Kadau}, \citenamefont {Schmitt},
  \citenamefont {Wenzel},\ and\ \citenamefont {Pfau}}]{Pfau:PRL:2016}%
  \BibitemOpen
  \bibfield  {author} {\bibinfo {author} {\bibfnamefont {Igor}\ \bibnamefont
  {Ferrier-Barbut}}, \bibinfo {author} {\bibfnamefont {Holger}\ \bibnamefont
  {Kadau}}, \bibinfo {author} {\bibfnamefont {Matthias}\ \bibnamefont
  {Schmitt}}, \bibinfo {author} {\bibfnamefont {Matthias}\ \bibnamefont
  {Wenzel}}, \ and\ \bibinfo {author} {\bibfnamefont {Tilman}\ \bibnamefont
  {Pfau}},\ }\bibfield  {title} {\enquote {\bibinfo {title} {Observation of
  quantum droplets in a strongly dipolar bose gas},}\ }\href {\doibase
  10.1103/PhysRevLett.116.215301} {\bibfield  {journal} {\bibinfo  {journal}
  {Phys. Rev. Lett.}\ }\textbf {\bibinfo {volume} {116}},\ \bibinfo {pages}
  {215301} (\bibinfo {year} {2016})}\BibitemShut {NoStop}%
\bibitem [{\citenamefont {Chomaz}\ \emph {et~al.}(2016)\citenamefont {Chomaz},
  \citenamefont {Baier}, \citenamefont {Petter}, \citenamefont {Mark},
  \citenamefont {W\"achtler}, \citenamefont {Santos},\ and\ \citenamefont
  {Ferlaino}}]{ferlaino16}%
  \BibitemOpen
  \bibfield  {author} {\bibinfo {author} {\bibfnamefont {L.}~\bibnamefont
  {Chomaz}}, \bibinfo {author} {\bibfnamefont {S.}~\bibnamefont {Baier}},
  \bibinfo {author} {\bibfnamefont {D.}~\bibnamefont {Petter}}, \bibinfo
  {author} {\bibfnamefont {M.~J.}\ \bibnamefont {Mark}}, \bibinfo {author}
  {\bibfnamefont {F.}~\bibnamefont {W\"achtler}}, \bibinfo {author}
  {\bibfnamefont {L.}~\bibnamefont {Santos}}, \ and\ \bibinfo {author}
  {\bibfnamefont {F.}~\bibnamefont {Ferlaino}},\ }\bibfield  {title} {\enquote
  {\bibinfo {title} {Quantum-fluctuation-driven crossover from a dilute
  bose-einstein condensate to a macrodroplet in a dipolar quantum fluid},}\
  }\href {\doibase 10.1103/PhysRevX.6.041039} {\bibfield  {journal} {\bibinfo
  {journal} {Phys. Rev. X}\ }\textbf {\bibinfo {volume} {6}},\ \bibinfo {pages}
  {041039} (\bibinfo {year} {2016})}\BibitemShut {NoStop}%
\bibitem [{\citenamefont {B\"ottcher}\ \emph
  {et~al.}(2019{\natexlab{a}})\citenamefont {B\"ottcher}, \citenamefont
  {Wenzel}, \citenamefont {Schmidt}, \citenamefont {Guo}, \citenamefont
  {Langen}, \citenamefont {Ferrier-Barbut}, \citenamefont {Pfau}, \citenamefont
  {Bomb\'{\i}n}, \citenamefont {S\'anchez-Baena}, \citenamefont {Boronat},\
  and\ \citenamefont {Mazzanti}}]{bottcher19}%
  \BibitemOpen
  \bibfield  {author} {\bibinfo {author} {\bibfnamefont {Fabian}\ \bibnamefont
  {B\"ottcher}}, \bibinfo {author} {\bibfnamefont {Matthias}\ \bibnamefont
  {Wenzel}}, \bibinfo {author} {\bibfnamefont {Jan-Niklas}\ \bibnamefont
  {Schmidt}}, \bibinfo {author} {\bibfnamefont {Mingyang}\ \bibnamefont {Guo}},
  \bibinfo {author} {\bibfnamefont {Tim}\ \bibnamefont {Langen}}, \bibinfo
  {author} {\bibfnamefont {Igor}\ \bibnamefont {Ferrier-Barbut}}, \bibinfo
  {author} {\bibfnamefont {Tilman}\ \bibnamefont {Pfau}}, \bibinfo {author}
  {\bibfnamefont {Ra\'ul}\ \bibnamefont {Bomb\'{\i}n}}, \bibinfo {author}
  {\bibfnamefont {Joan}\ \bibnamefont {S\'anchez-Baena}}, \bibinfo {author}
  {\bibfnamefont {Jordi}\ \bibnamefont {Boronat}}, \ and\ \bibinfo {author}
  {\bibfnamefont {Ferran}\ \bibnamefont {Mazzanti}},\ }\bibfield  {title}
  {\enquote {\bibinfo {title} {Dilute dipolar quantum droplets beyond the
  extended gross-pitaevskii equation},}\ }\href {\doibase
  10.1103/PhysRevResearch.1.033088} {\bibfield  {journal} {\bibinfo  {journal}
  {Phys. Rev. Res.}\ }\textbf {\bibinfo {volume} {1}},\ \bibinfo {pages}
  {033088} (\bibinfo {year} {2019}{\natexlab{a}})}\BibitemShut {NoStop}%
\bibitem [{\citenamefont {Tanzi}\ \emph
  {et~al.}(2019{\natexlab{a}})\citenamefont {Tanzi}, \citenamefont {Lucioni},
  \citenamefont {Fam\`a}, \citenamefont {Catani}, \citenamefont {Fioretti},
  \citenamefont {Gabbanini}, \citenamefont {Bisset}, \citenamefont {Santos},\
  and\ \citenamefont {Modugno}}]{Modugno:PRL:2019}%
  \BibitemOpen
  \bibfield  {author} {\bibinfo {author} {\bibfnamefont {L.}~\bibnamefont
  {Tanzi}}, \bibinfo {author} {\bibfnamefont {E.}~\bibnamefont {Lucioni}},
  \bibinfo {author} {\bibfnamefont {F.}~\bibnamefont {Fam\`a}}, \bibinfo
  {author} {\bibfnamefont {J.}~\bibnamefont {Catani}}, \bibinfo {author}
  {\bibfnamefont {A.}~\bibnamefont {Fioretti}}, \bibinfo {author}
  {\bibfnamefont {C.}~\bibnamefont {Gabbanini}}, \bibinfo {author}
  {\bibfnamefont {R.~N.}\ \bibnamefont {Bisset}}, \bibinfo {author}
  {\bibfnamefont {L.}~\bibnamefont {Santos}}, \ and\ \bibinfo {author}
  {\bibfnamefont {G.}~\bibnamefont {Modugno}},\ }\bibfield  {title} {\enquote
  {\bibinfo {title} {Observation of a dipolar quantum gas with metastable
  supersolid properties},}\ }\href {\doibase 10.1103/PhysRevLett.122.130405}
  {\bibfield  {journal} {\bibinfo  {journal} {Phys. Rev. Lett.}\ }\textbf
  {\bibinfo {volume} {122}},\ \bibinfo {pages} {130405} (\bibinfo {year}
  {2019}{\natexlab{a}})}\BibitemShut {NoStop}%
\bibitem [{\citenamefont {B\"ottcher}\ \emph
  {et~al.}(2019{\natexlab{b}})\citenamefont {B\"ottcher}, \citenamefont
  {Schmidt}, \citenamefont {Wenzel}, \citenamefont {Hertkorn}, \citenamefont
  {Guo}, \citenamefont {Langen},\ and\ \citenamefont {Pfau}}]{Pfau:PRX:2019}%
  \BibitemOpen
  \bibfield  {author} {\bibinfo {author} {\bibfnamefont {Fabian}\ \bibnamefont
  {B\"ottcher}}, \bibinfo {author} {\bibfnamefont {Jan-Niklas}\ \bibnamefont
  {Schmidt}}, \bibinfo {author} {\bibfnamefont {Matthias}\ \bibnamefont
  {Wenzel}}, \bibinfo {author} {\bibfnamefont {Jens}\ \bibnamefont {Hertkorn}},
  \bibinfo {author} {\bibfnamefont {Mingyang}\ \bibnamefont {Guo}}, \bibinfo
  {author} {\bibfnamefont {Tim}\ \bibnamefont {Langen}}, \ and\ \bibinfo
  {author} {\bibfnamefont {Tilman}\ \bibnamefont {Pfau}},\ }\bibfield  {title}
  {\enquote {\bibinfo {title} {Transient supersolid properties in an array of
  dipolar quantum droplets},}\ }\href {\doibase 10.1103/PhysRevX.9.011051}
  {\bibfield  {journal} {\bibinfo  {journal} {Phys. Rev. X}\ }\textbf {\bibinfo
  {volume} {9}},\ \bibinfo {pages} {011051} (\bibinfo {year}
  {2019}{\natexlab{b}})}\BibitemShut {NoStop}%
\bibitem [{\citenamefont {Chomaz}\ \emph {et~al.}(2019)\citenamefont {Chomaz},
  \citenamefont {Petter}, \citenamefont {Ilzh\"ofer}, \citenamefont {Natale},
  \citenamefont {Trautmann}, \citenamefont {Politi}, \citenamefont
  {Durastante}, \citenamefont {van Bijnen}, \citenamefont {Patscheider},
  \citenamefont {Sohmen}, \citenamefont {Mark},\ and\ \citenamefont
  {Ferlaino}}]{Ferlaino:PRX:2019}%
  \BibitemOpen
  \bibfield  {author} {\bibinfo {author} {\bibfnamefont {L.}~\bibnamefont
  {Chomaz}}, \bibinfo {author} {\bibfnamefont {D.}~\bibnamefont {Petter}},
  \bibinfo {author} {\bibfnamefont {P.}~\bibnamefont {Ilzh\"ofer}}, \bibinfo
  {author} {\bibfnamefont {G.}~\bibnamefont {Natale}}, \bibinfo {author}
  {\bibfnamefont {A.}~\bibnamefont {Trautmann}}, \bibinfo {author}
  {\bibfnamefont {C.}~\bibnamefont {Politi}}, \bibinfo {author} {\bibfnamefont
  {G.}~\bibnamefont {Durastante}}, \bibinfo {author} {\bibfnamefont {R.~M.~W.}\
  \bibnamefont {van Bijnen}}, \bibinfo {author} {\bibfnamefont
  {A.}~\bibnamefont {Patscheider}}, \bibinfo {author} {\bibfnamefont
  {M.}~\bibnamefont {Sohmen}}, \bibinfo {author} {\bibfnamefont {M.~J.}\
  \bibnamefont {Mark}}, \ and\ \bibinfo {author} {\bibfnamefont
  {F.}~\bibnamefont {Ferlaino}},\ }\bibfield  {title} {\enquote {\bibinfo
  {title} {Long-lived and transient supersolid behaviors in dipolar quantum
  gases},}\ }\href {\doibase 10.1103/PhysRevX.9.021012} {\bibfield  {journal}
  {\bibinfo  {journal} {Phys. Rev. X}\ }\textbf {\bibinfo {volume} {9}},\
  \bibinfo {pages} {021012} (\bibinfo {year} {2019})}\BibitemShut {NoStop}%
\bibitem [{\citenamefont {Tanzi}\ \emph
  {et~al.}(2019{\natexlab{b}})\citenamefont {Tanzi}, \citenamefont {Roccuzzo},
  \citenamefont {Lucioni}, \citenamefont {Fam{\`a}}, \citenamefont {Fioretti},
  \citenamefont {Gabbanini}, \citenamefont {Modugno}, \citenamefont {Recati},\
  and\ \citenamefont {Stringari}}]{Tanzi:Nature:2019}%
  \BibitemOpen
  \bibfield  {author} {\bibinfo {author} {\bibfnamefont {L.}~\bibnamefont
  {Tanzi}}, \bibinfo {author} {\bibfnamefont {S.~M.}\ \bibnamefont {Roccuzzo}},
  \bibinfo {author} {\bibfnamefont {E.}~\bibnamefont {Lucioni}}, \bibinfo
  {author} {\bibfnamefont {F.}~\bibnamefont {Fam{\`a}}}, \bibinfo {author}
  {\bibfnamefont {A.}~\bibnamefont {Fioretti}}, \bibinfo {author}
  {\bibfnamefont {C.}~\bibnamefont {Gabbanini}}, \bibinfo {author}
  {\bibfnamefont {G.}~\bibnamefont {Modugno}}, \bibinfo {author} {\bibfnamefont
  {A.}~\bibnamefont {Recati}}, \ and\ \bibinfo {author} {\bibfnamefont
  {S.}~\bibnamefont {Stringari}},\ }\bibfield  {title} {\enquote {\bibinfo
  {title} {Supersolid symmetry breaking from compressional oscillations in a
  dipolar quantum gas},}\ }\href {\doibase 10.1038/s41586-019-1568-6}
  {\bibfield  {journal} {\bibinfo  {journal} {Nature}\ }\textbf {\bibinfo
  {volume} {574}},\ \bibinfo {pages} {382--385} (\bibinfo {year}
  {2019}{\natexlab{b}})}\BibitemShut {NoStop}%
\bibitem [{\citenamefont {Guo}\ \emph {et~al.}(2019)\citenamefont {Guo},
  \citenamefont {B{\"o}ttcher}, \citenamefont {Hertkorn}, \citenamefont
  {Schmidt}, \citenamefont {Wenzel}, \citenamefont {B{\"u}chler}, \citenamefont
  {Langen},\ and\ \citenamefont {Pfau}}]{Guo:Nature:2019}%
  \BibitemOpen
  \bibfield  {author} {\bibinfo {author} {\bibfnamefont {Mingyang}\
  \bibnamefont {Guo}}, \bibinfo {author} {\bibfnamefont {Fabian}\ \bibnamefont
  {B{\"o}ttcher}}, \bibinfo {author} {\bibfnamefont {Jens}\ \bibnamefont
  {Hertkorn}}, \bibinfo {author} {\bibfnamefont {Jan-Niklas}\ \bibnamefont
  {Schmidt}}, \bibinfo {author} {\bibfnamefont {Matthias}\ \bibnamefont
  {Wenzel}}, \bibinfo {author} {\bibfnamefont {Hans~Peter}\ \bibnamefont
  {B{\"u}chler}}, \bibinfo {author} {\bibfnamefont {Tim}\ \bibnamefont
  {Langen}}, \ and\ \bibinfo {author} {\bibfnamefont {Tilman}\ \bibnamefont
  {Pfau}},\ }\bibfield  {title} {\enquote {\bibinfo {title} {The low-energy
  goldstone mode in a trapped dipolar supersolid},}\ }\href {\doibase
  10.1038/s41586-019-1569-5} {\bibfield  {journal} {\bibinfo  {journal}
  {Nature}\ }\textbf {\bibinfo {volume} {574}},\ \bibinfo {pages} {386--389}
  (\bibinfo {year} {2019})}\BibitemShut {NoStop}%
\bibitem [{\citenamefont {Tanzi}\ \emph {et~al.}(2021)\citenamefont {Tanzi},
  \citenamefont {Maloberti}, \citenamefont {Biagioni}, \citenamefont
  {Fioretti}, \citenamefont {Gabbanini},\ and\ \citenamefont
  {Modugno}}]{Tanzi:Science:2021}%
  \BibitemOpen
  \bibfield  {author} {\bibinfo {author} {\bibfnamefont {L.}~\bibnamefont
  {Tanzi}}, \bibinfo {author} {\bibfnamefont {J.~G.}\ \bibnamefont
  {Maloberti}}, \bibinfo {author} {\bibfnamefont {G.}~\bibnamefont {Biagioni}},
  \bibinfo {author} {\bibfnamefont {A.}~\bibnamefont {Fioretti}}, \bibinfo
  {author} {\bibfnamefont {C.}~\bibnamefont {Gabbanini}}, \ and\ \bibinfo
  {author} {\bibfnamefont {G.}~\bibnamefont {Modugno}},\ }\bibfield  {title}
  {\enquote {\bibinfo {title} {Evidence of superfluidity in a dipolar
  supersolid from nonclassical rotational inertia},}\ }\href {\doibase
  10.1126/science.aba4309} {\bibfield  {journal} {\bibinfo  {journal}
  {Science}\ }\textbf {\bibinfo {volume} {371}},\ \bibinfo {pages} {1162--1165}
  (\bibinfo {year} {2021})}\BibitemShut {NoStop}%
\bibitem [{\citenamefont {Norcia}\ \emph {et~al.}(2021)\citenamefont {Norcia},
  \citenamefont {Politi}, \citenamefont {Klaus}, \citenamefont {Poli},
  \citenamefont {Sohmen}, \citenamefont {Mark}, \citenamefont {Bisset},
  \citenamefont {Santos},\ and\ \citenamefont {Ferlaino}}]{norcia21:nature}%
  \BibitemOpen
  \bibfield  {author} {\bibinfo {author} {\bibfnamefont {Matthew~A.}\
  \bibnamefont {Norcia}}, \bibinfo {author} {\bibfnamefont {Claudia}\
  \bibnamefont {Politi}}, \bibinfo {author} {\bibfnamefont {Lauritz}\
  \bibnamefont {Klaus}}, \bibinfo {author} {\bibfnamefont {Elena}\ \bibnamefont
  {Poli}}, \bibinfo {author} {\bibfnamefont {Maximilian}\ \bibnamefont
  {Sohmen}}, \bibinfo {author} {\bibfnamefont {Manfred~J.}\ \bibnamefont
  {Mark}}, \bibinfo {author} {\bibfnamefont {Russell~N.}\ \bibnamefont
  {Bisset}}, \bibinfo {author} {\bibfnamefont {Luis}\ \bibnamefont {Santos}}, \
  and\ \bibinfo {author} {\bibfnamefont {Francesca}\ \bibnamefont {Ferlaino}},\
  }\bibfield  {title} {\enquote {\bibinfo {title} {Two-dimensional
  supersolidity in a dipolar quantum gas},}\ }\href {\doibase
  10.1038/s41586-021-03725-7} {\bibfield  {journal} {\bibinfo  {journal}
  {Nature}\ }\textbf {\bibinfo {volume} {596}},\ \bibinfo {pages} {357--361}
  (\bibinfo {year} {2021})}\BibitemShut {NoStop}%
\bibitem [{\citenamefont {Biagioni}\ \emph {et~al.}(2022)\citenamefont
  {Biagioni}, \citenamefont {Antolini}, \citenamefont {Ala\~na}, \citenamefont
  {Modugno}, \citenamefont {Fioretti}, \citenamefont {Gabbanini}, \citenamefont
  {Tanzi},\ and\ \citenamefont {Modugno}}]{BiagioniPRX2022}%
  \BibitemOpen
  \bibfield  {author} {\bibinfo {author} {\bibfnamefont {Giulio}\ \bibnamefont
  {Biagioni}}, \bibinfo {author} {\bibfnamefont {Nicol\`o}\ \bibnamefont
  {Antolini}}, \bibinfo {author} {\bibfnamefont {Aitor}\ \bibnamefont
  {Ala\~na}}, \bibinfo {author} {\bibfnamefont {Michele}\ \bibnamefont
  {Modugno}}, \bibinfo {author} {\bibfnamefont {Andrea}\ \bibnamefont
  {Fioretti}}, \bibinfo {author} {\bibfnamefont {Carlo}\ \bibnamefont
  {Gabbanini}}, \bibinfo {author} {\bibfnamefont {Luca}\ \bibnamefont {Tanzi}},
  \ and\ \bibinfo {author} {\bibfnamefont {Giovanni}\ \bibnamefont {Modugno}},\
  }\bibfield  {title} {\enquote {\bibinfo {title} {Dimensional crossover in the
  superfluid-supersolid quantum phase transition},}\ }\href {\doibase
  10.1103/PhysRevX.12.021019} {\bibfield  {journal} {\bibinfo  {journal} {Phys.
  Rev. X}\ }\textbf {\bibinfo {volume} {12}},\ \bibinfo {pages} {021019}
  (\bibinfo {year} {2022})}\BibitemShut {NoStop}%
\bibitem [{\citenamefont {Sohmen}\ \emph {et~al.}(2021)\citenamefont {Sohmen},
  \citenamefont {Politi}, \citenamefont {Klaus}, \citenamefont {Chomaz},
  \citenamefont {Mark}, \citenamefont {Norcia},\ and\ \citenamefont
  {Ferlaino}}]{Ferlaino:PRL:2021}%
  \BibitemOpen
  \bibfield  {author} {\bibinfo {author} {\bibfnamefont {Maximilian}\
  \bibnamefont {Sohmen}}, \bibinfo {author} {\bibfnamefont {Claudia}\
  \bibnamefont {Politi}}, \bibinfo {author} {\bibfnamefont {Lauritz}\
  \bibnamefont {Klaus}}, \bibinfo {author} {\bibfnamefont {Lauriane}\
  \bibnamefont {Chomaz}}, \bibinfo {author} {\bibfnamefont {Manfred~J.}\
  \bibnamefont {Mark}}, \bibinfo {author} {\bibfnamefont {Matthew~A.}\
  \bibnamefont {Norcia}}, \ and\ \bibinfo {author} {\bibfnamefont {Francesca}\
  \bibnamefont {Ferlaino}},\ }\bibfield  {title} {\enquote {\bibinfo {title}
  {Birth, life, and death of a dipolar supersolid},}\ }\href {\doibase
  10.1103/PhysRevLett.126.233401} {\bibfield  {journal} {\bibinfo  {journal}
  {Phys. Rev. Lett.}\ }\textbf {\bibinfo {volume} {126}},\ \bibinfo {pages}
  {233401} (\bibinfo {year} {2021})}\BibitemShut {NoStop}%
\bibitem [{\citenamefont {S{\'a}nchez-Baena}\ \emph {et~al.}(2023)\citenamefont
  {S{\'a}nchez-Baena}, \citenamefont {Politi}, \citenamefont {Maucher},
  \citenamefont {Ferlaino},\ and\ \citenamefont {Pohl}}]{baena22}%
  \BibitemOpen
  \bibfield  {author} {\bibinfo {author} {\bibfnamefont {J.}~\bibnamefont
  {S{\'a}nchez-Baena}}, \bibinfo {author} {\bibfnamefont {C.}~\bibnamefont
  {Politi}}, \bibinfo {author} {\bibfnamefont {F.}~\bibnamefont {Maucher}},
  \bibinfo {author} {\bibfnamefont {F.}~\bibnamefont {Ferlaino}}, \ and\
  \bibinfo {author} {\bibfnamefont {T.}~\bibnamefont {Pohl}},\ }\bibfield
  {title} {\enquote {\bibinfo {title} {Heating a dipolar quantum fluid into a
  solid},}\ }\href {\doibase 10.1038/s41467-023-37207-3} {\bibfield  {journal}
  {\bibinfo  {journal} {Nature Communications}\ }\textbf {\bibinfo {volume}
  {14}},\ \bibinfo {pages} {1868} (\bibinfo {year} {2023})}\BibitemShut
  {NoStop}%
\bibitem [{\citenamefont {S\'anchez-Baena}\ \emph {et~al.}(2024)\citenamefont
  {S\'anchez-Baena}, \citenamefont {Pohl},\ and\ \citenamefont
  {Maucher}}]{baena24}%
  \BibitemOpen
  \bibfield  {author} {\bibinfo {author} {\bibfnamefont {J.}~\bibnamefont
  {S\'anchez-Baena}}, \bibinfo {author} {\bibfnamefont {T.}~\bibnamefont
  {Pohl}}, \ and\ \bibinfo {author} {\bibfnamefont {F.}~\bibnamefont
  {Maucher}},\ }\bibfield  {title} {\enquote {\bibinfo {title}
  {Superfluid-supersolid phase transition of elongated dipolar bose-einstein
  condensates at finite temperatures},}\ }\href {\doibase
  10.1103/PhysRevResearch.6.023183} {\bibfield  {journal} {\bibinfo  {journal}
  {Phys. Rev. Res.}\ }\textbf {\bibinfo {volume} {6}},\ \bibinfo {pages}
  {023183} (\bibinfo {year} {2024})}\BibitemShut {NoStop}%
\bibitem [{\citenamefont {He}\ \emph {et~al.}(2025)\citenamefont {He},
  \citenamefont {S\'anchez-Baena}, \citenamefont {Maucher},\ and\ \citenamefont
  {Zhang}}]{He2024}%
  \BibitemOpen
  \bibfield  {author} {\bibinfo {author} {\bibfnamefont {Liang-Jun}\
  \bibnamefont {He}}, \bibinfo {author} {\bibfnamefont {Juan}\ \bibnamefont
  {S\'anchez-Baena}}, \bibinfo {author} {\bibfnamefont {Fabian}\ \bibnamefont
  {Maucher}}, \ and\ \bibinfo {author} {\bibfnamefont {Yong-Chang}\
  \bibnamefont {Zhang}},\ }\bibfield  {title} {\enquote {\bibinfo {title}
  {Accessing elusive two-dimensional phases of dipolar bose-einstein
  condensates by finite temperature},}\ }\href {\doibase
  10.1103/PhysRevResearch.7.023019} {\bibfield  {journal} {\bibinfo  {journal}
  {Phys. Rev. Res.}\ }\textbf {\bibinfo {volume} {7}},\ \bibinfo {pages}
  {023019} (\bibinfo {year} {2025})}\BibitemShut {NoStop}%
\bibitem [{\citenamefont {Lu}\ \emph {et~al.}(2011)\citenamefont {Lu},
  \citenamefont {Burdick}, \citenamefont {Youn},\ and\ \citenamefont
  {Lev}}]{Mingwu:PRL:2011}%
  \BibitemOpen
  \bibfield  {author} {\bibinfo {author} {\bibfnamefont {Mingwu}\ \bibnamefont
  {Lu}}, \bibinfo {author} {\bibfnamefont {Nathaniel~Q.}\ \bibnamefont
  {Burdick}}, \bibinfo {author} {\bibfnamefont {Seo~Ho}\ \bibnamefont {Youn}},
  \ and\ \bibinfo {author} {\bibfnamefont {Benjamin~L.}\ \bibnamefont {Lev}},\
  }\bibfield  {title} {\enquote {\bibinfo {title} {Strongly dipolar
  bose-einstein condensate of dysprosium},}\ }\href {\doibase
  10.1103/PhysRevLett.107.190401} {\bibfield  {journal} {\bibinfo  {journal}
  {Phys. Rev. Lett.}\ }\textbf {\bibinfo {volume} {107}},\ \bibinfo {pages}
  {190401} (\bibinfo {year} {2011})}\BibitemShut {NoStop}%
\bibitem [{\citenamefont {Aikawa}\ \emph {et~al.}(2012)\citenamefont {Aikawa},
  \citenamefont {Frisch}, \citenamefont {Mark}, \citenamefont {Baier},
  \citenamefont {Rietzler}, \citenamefont {Grimm},\ and\ \citenamefont
  {Ferlaino}}]{Aikawa:PRL:2012}%
  \BibitemOpen
  \bibfield  {author} {\bibinfo {author} {\bibfnamefont {K.}~\bibnamefont
  {Aikawa}}, \bibinfo {author} {\bibfnamefont {A.}~\bibnamefont {Frisch}},
  \bibinfo {author} {\bibfnamefont {M.}~\bibnamefont {Mark}}, \bibinfo {author}
  {\bibfnamefont {S.}~\bibnamefont {Baier}}, \bibinfo {author} {\bibfnamefont
  {A.}~\bibnamefont {Rietzler}}, \bibinfo {author} {\bibfnamefont
  {R.}~\bibnamefont {Grimm}}, \ and\ \bibinfo {author} {\bibfnamefont
  {F.}~\bibnamefont {Ferlaino}},\ }\bibfield  {title} {\enquote {\bibinfo
  {title} {Bose-einstein condensation of erbium},}\ }\href {\doibase
  10.1103/PhysRevLett.108.210401} {\bibfield  {journal} {\bibinfo  {journal}
  {Phys. Rev. Lett.}\ }\textbf {\bibinfo {volume} {108}},\ \bibinfo {pages}
  {210401} (\bibinfo {year} {2012})}\BibitemShut {NoStop}%
\bibitem [{\citenamefont {Giorgini}\ \emph {et~al.}(1997)\citenamefont
  {Giorgini}, \citenamefont {Pitaevskii},\ and\ \citenamefont
  {Stringari}}]{Giorgini:JLTP:1997}%
  \BibitemOpen
  \bibfield  {author} {\bibinfo {author} {\bibfnamefont {S.}~\bibnamefont
  {Giorgini}}, \bibinfo {author} {\bibfnamefont {L.~P.}\ \bibnamefont
  {Pitaevskii}}, \ and\ \bibinfo {author} {\bibfnamefont {S.}~\bibnamefont
  {Stringari}},\ }\bibfield  {title} {\enquote {\bibinfo {title}
  {Thermodynamics of a trapped bose-condensed gas},}\ }\href {\doibase
  10.1007/s10909-005-0089-x} {\bibfield  {journal} {\bibinfo  {journal}
  {Journal of Low Temperature Physics}\ }\textbf {\bibinfo {volume} {109}},\
  \bibinfo {pages} {309--355} (\bibinfo {year} {1997})}\BibitemShut {NoStop}%
\bibitem [{\citenamefont {Boudjemâa}(2017)}]{boudjemaa:aop:2017}%
  \BibitemOpen
  \bibfield  {author} {\bibinfo {author} {\bibfnamefont {Abdelâali}\
  \bibnamefont {Boudjemâa}},\ }\bibfield  {title} {\enquote {\bibinfo {title}
  {Quantum dilute droplets of dipolar bosons at finite temperature},}\ }\href
  {\doibase https://doi.org/10.1016/j.aop.2017.03.020} {\bibfield  {journal}
  {\bibinfo  {journal} {Annals of Physics}\ }\textbf {\bibinfo {volume}
  {381}},\ \bibinfo {pages} {68--79} (\bibinfo {year} {2017})}\BibitemShut
  {NoStop}%
\bibitem [{\citenamefont {Boudjem\^aa}(2018)}]{boudjemaa:pra:2018}%
  \BibitemOpen
  \bibfield  {author} {\bibinfo {author} {\bibfnamefont {Abdel\^aali}\
  \bibnamefont {Boudjem\^aa}},\ }\bibfield  {title} {\enquote {\bibinfo {title}
  {Fluctuations and quantum self-bound droplets in a dipolar bose-bose
  mixture},}\ }\href {\doibase 10.1103/PhysRevA.98.033612} {\bibfield
  {journal} {\bibinfo  {journal} {Phys. Rev. A}\ }\textbf {\bibinfo {volume}
  {98}},\ \bibinfo {pages} {033612} (\bibinfo {year} {2018})}\BibitemShut
  {NoStop}%
\bibitem [{\citenamefont {Aybar}\ and\ \citenamefont {Oktel}(2019)}]{oktel19}%
  \BibitemOpen
  \bibfield  {author} {\bibinfo {author} {\bibfnamefont {E.}~\bibnamefont
  {Aybar}}\ and\ \bibinfo {author} {\bibfnamefont {M.~\"O.}\ \bibnamefont
  {Oktel}},\ }\bibfield  {title} {\enquote {\bibinfo {title}
  {Temperature-dependent density profiles of dipolar droplets},}\ }\href
  {\doibase 10.1103/PhysRevA.99.013620} {\bibfield  {journal} {\bibinfo
  {journal} {Phys. Rev. A}\ }\textbf {\bibinfo {volume} {99}},\ \bibinfo
  {pages} {013620} (\bibinfo {year} {2019})}\BibitemShut {NoStop}%
\bibitem [{\citenamefont {\"Ozt\"urk}\ \emph {et~al.}(2020)\citenamefont
  {\"Ozt\"urk}, \citenamefont {Aybar},\ and\ \citenamefont {Oktel}}]{oktel20}%
  \BibitemOpen
  \bibfield  {author} {\bibinfo {author} {\bibfnamefont {S.~F.}\ \bibnamefont
  {\"Ozt\"urk}}, \bibinfo {author} {\bibfnamefont {Enes}\ \bibnamefont
  {Aybar}}, \ and\ \bibinfo {author} {\bibfnamefont {M.~\"O.}\ \bibnamefont
  {Oktel}},\ }\bibfield  {title} {\enquote {\bibinfo {title} {Temperature
  dependence of the density and excitations of dipolar droplets},}\ }\href
  {\doibase 10.1103/PhysRevA.102.033329} {\bibfield  {journal} {\bibinfo
  {journal} {Phys. Rev. A}\ }\textbf {\bibinfo {volume} {102}},\ \bibinfo
  {pages} {033329} (\bibinfo {year} {2020})}\BibitemShut {NoStop}%
\bibitem [{\citenamefont {Lima}\ and\ \citenamefont
  {Pelster}(2011)}]{Lima:2011eq}%
  \BibitemOpen
  \bibfield  {author} {\bibinfo {author} {\bibfnamefont {Aristeu R~P}\
  \bibnamefont {Lima}}\ and\ \bibinfo {author} {\bibfnamefont {Axel}\
  \bibnamefont {Pelster}},\ }\bibfield  {title} {\enquote {\bibinfo {title}
  {{Quantum fluctuations in dipolar Bose gases}},}\ }\href@noop {} {\bibfield
  {journal} {\bibinfo  {journal} {Physical Review A}\ }\textbf {\bibinfo
  {volume} {84}},\ \bibinfo {pages} {041604--4} (\bibinfo {year}
  {2011})}\BibitemShut {NoStop}%
\bibitem [{\citenamefont {Lima}\ and\ \citenamefont
  {Pelster}(2012)}]{pelster12}%
  \BibitemOpen
  \bibfield  {author} {\bibinfo {author} {\bibfnamefont {A.~R.~P.}\
  \bibnamefont {Lima}}\ and\ \bibinfo {author} {\bibfnamefont {A.}~\bibnamefont
  {Pelster}},\ }\bibfield  {title} {\enquote {\bibinfo {title} {Beyond
  mean-field low-lying excitations of dipolar bose gases},}\ }\href {\doibase
  10.1103/PhysRevA.86.063609} {\bibfield  {journal} {\bibinfo  {journal} {Phys.
  Rev. A}\ }\textbf {\bibinfo {volume} {86}},\ \bibinfo {pages} {063609}
  (\bibinfo {year} {2012})}\BibitemShut {NoStop}%
\bibitem [{\citenamefont {Dalfovo}\ \emph {et~al.}(1999)\citenamefont
  {Dalfovo}, \citenamefont {Giorgini}, \citenamefont {Pitaevskii},\ and\
  \citenamefont {Stringari}}]{dalfovo99}%
  \BibitemOpen
  \bibfield  {author} {\bibinfo {author} {\bibfnamefont {Franco}\ \bibnamefont
  {Dalfovo}}, \bibinfo {author} {\bibfnamefont {Stefano}\ \bibnamefont
  {Giorgini}}, \bibinfo {author} {\bibfnamefont {Lev~P.}\ \bibnamefont
  {Pitaevskii}}, \ and\ \bibinfo {author} {\bibfnamefont {Sandro}\ \bibnamefont
  {Stringari}},\ }\bibfield  {title} {\enquote {\bibinfo {title} {Theory of
  bose-einstein condensation in trapped gases},}\ }\href {\doibase
  10.1103/RevModPhys.71.463} {\bibfield  {journal} {\bibinfo  {journal} {Rev.
  Mod. Phys.}\ }\textbf {\bibinfo {volume} {71}},\ \bibinfo {pages} {463--512}
  (\bibinfo {year} {1999})}\BibitemShut {NoStop}%
\bibitem [{\citenamefont {Karman}\ \emph {et~al.}(2025)\citenamefont {Karman},
  \citenamefont {Bigagli}, \citenamefont {Yuan}, \citenamefont {Zhang},
  \citenamefont {Stevenson},\ and\ \citenamefont {Will}}]{karman2025:arxiv}%
  \BibitemOpen
  \bibfield  {author} {\bibinfo {author} {\bibfnamefont {Tijs}\ \bibnamefont
  {Karman}}, \bibinfo {author} {\bibfnamefont {Niccolò}\ \bibnamefont
  {Bigagli}}, \bibinfo {author} {\bibfnamefont {Weijun}\ \bibnamefont {Yuan}},
  \bibinfo {author} {\bibfnamefont {Siwei}\ \bibnamefont {Zhang}}, \bibinfo
  {author} {\bibfnamefont {Ian}\ \bibnamefont {Stevenson}}, \ and\ \bibinfo
  {author} {\bibfnamefont {Sebastian}\ \bibnamefont {Will}},\ }\href
  {https://arxiv.org/abs/2501.08095} {\enquote {\bibinfo {title} {Double
  microwave shielding},}\ } (\bibinfo {year} {2025}),\ \Eprint
  {http://arxiv.org/abs/2501.08095} {arXiv:2501.08095 [cond-mat.quant-gas]}
  \BibitemShut {NoStop}%
\bibitem [{\citenamefont {Deng}\ \emph {et~al.}(2025)\citenamefont {Deng},
  \citenamefont {Hu}, \citenamefont {Jin}, \citenamefont {Yi},\ and\
  \citenamefont {Shi}}]{deng2025:arxiv}%
  \BibitemOpen
  \bibfield  {author} {\bibinfo {author} {\bibfnamefont {Fulin}\ \bibnamefont
  {Deng}}, \bibinfo {author} {\bibfnamefont {Xinyuan}\ \bibnamefont {Hu}},
  \bibinfo {author} {\bibfnamefont {Wei-Jian}\ \bibnamefont {Jin}}, \bibinfo
  {author} {\bibfnamefont {Su}~\bibnamefont {Yi}}, \ and\ \bibinfo {author}
  {\bibfnamefont {Tao}\ \bibnamefont {Shi}},\ }\href
  {https://arxiv.org/abs/2501.05210} {\enquote {\bibinfo {title} {Two- and
  many-body physics of ultracold molecules dressed by dual microwave fields},}\
  } (\bibinfo {year} {2025}),\ \Eprint {http://arxiv.org/abs/2501.05210}
  {arXiv:2501.05210 [cond-mat.quant-gas]} \BibitemShut {NoStop}%
\bibitem [{\citenamefont {Boudjem\^aa}(2025)}]{boudjemaa:pra:2025}%
  \BibitemOpen
  \bibfield  {author} {\bibinfo {author} {\bibfnamefont {Abdel\^aali}\
  \bibnamefont {Boudjem\^aa}},\ }\bibfield  {title} {\enquote {\bibinfo {title}
  {Nonequilibrium quench dynamics of bose-einstein condensates of
  microwave-shielded polar molecules},}\ }\href {\doibase 10.1103/6b82-sb2j}
  {\bibfield  {journal} {\bibinfo  {journal} {Phys. Rev. A}\ }\textbf {\bibinfo
  {volume} {111}},\ \bibinfo {pages} {063315} (\bibinfo {year}
  {2025})}\BibitemShut {NoStop}%
\bibitem [{\citenamefont {Yi}\ and\ \citenamefont {You}(2001)}]{Yi2001}%
  \BibitemOpen
  \bibfield  {author} {\bibinfo {author} {\bibfnamefont {S.}~\bibnamefont
  {Yi}}\ and\ \bibinfo {author} {\bibfnamefont {L.}~\bibnamefont {You}},\
  }\bibfield  {title} {\enquote {\bibinfo {title} {Trapped condensates of atoms
  with dipole interactions},}\ }\href {\doibase 10.1103/PhysRevA.63.053607}
  {\bibfield  {journal} {\bibinfo  {journal} {Phys. Rev. A}\ }\textbf {\bibinfo
  {volume} {63}},\ \bibinfo {pages} {053607} (\bibinfo {year}
  {2001})}\BibitemShut {NoStop}%
\end{thebibliography}%

\end{document}